\documentclass[structabstract]{aa}  
\usepackage{graphicx}
\usepackage{txfonts}
\usepackage{calc}

\newcommand{\kms}{km\,s$^{-1}$}

\begin{document}
   \title{Magnitude-range brightness variations of overactive K giants}

   \author{K. Ol\'ah\inst{1}, A. Mo\'or\inst{1}, Zs. K\H ov\'ari\inst{1}, T. Granzer \inst{2}, K.G. Strassmeier\inst{2}, L. Kriskovics \inst{1}                \and 
               K. Vida\inst{1}
                          %\fnmsep\thanks{Just to show the usage of the elements in the author field}
          }

   \institute{Konkoly Observatory MTA CsFK, Konkoly Thege M. u. 15/17, 1121 Budapest, Hungary\\
              \email{olah@konkoly.hu}
         \and
             Leibniz Institute for Astrophysics Potsdam (AIP), An der Sternwarte 16, 14482 Potsdam, Germany
             }

   \date{Received ..., 2014; accepted ...}

  \abstract
   {Decades-long, phase-resolved photometry of overactive spotted cool stars has revealed that their long-term peak-to-peak light variations can be as large as one magnitude. Such brightness variations are too large to be solely explained by rotational modulation and/or a cyclic, or pseudo-cyclic, waxing and waning of surface spots and faculae as we see in the Sun. }
   %aims heading (mandatory)
   {We study three representative, overactive spotted K giants (IL\,Hya, XX\,Tri, and DM\,UMa) known to exhibit $V$-band light variations between 0\fm65--1\fm05. Our aim is to find the origin of their large brightness variation.}
   {We employ long-term phase-resolved multicolor photometry, mostly from automatic telescopes, covering 42\,yr for IL\,Hya, 28\,yr for XX\,Tri, and 34\,yr for DM\,UMa. For one target, IL\,Hya, we present a new Doppler image from NSO data taken in late 1996. Effective temperatures for our targets are determined from all well-sampled observing epochs and are based on a $V-I_C$ color-index calibration. }
   {The effective temperature change between the extrema of the rotational modulation for IL~Hya and XX~Tri is in the range 50--200~K. The bolometric flux during maximum of the rotational modulation, i.e., the least spotted states, varied by up to 39\%\ in IL\,Hya and up to 54\%\ in XX\,Tri over the course of our observations. We emphasize that for IL\,Hya this is just about half of the total luminosity variation that can be explained by the photospheric temperature (spots/faculae) changes, while for XX\,Tri it is even about one third. The long-term, 0\fm6 $V$-band variation of DM\,UMa is more difficult to explain because little or no $B-V$ color index change is observed on the same timescale. Placing the three stars with their light and color variations into H-R diagrams, we find that their overall luminosities are generally too low compared to predictions from current evolutionary tracks. }
  % conclusions heading (optional) 
   {A change in the stellar radius due to strong and variable magnetic fields during activity cycles likely plays a role in explaining the anomalous brightness and luminosity of our three targets. At least for IL\,Hya, a radius change of about 9\%
is suggested from $m_{\rm bol}$ and $T_{\rm eff}$, and is supported by independent $v\sin{i}$ measurements.
}

   \authorrunning{Ol\'ah et al.}
   \maketitle
%
%________________________________________________________________

\section{Introduction and target stars}\label{intro}

Magnetically active stars, including the Sun, are known to show complex, cyclic brightness variability with a large range of timescales of up to several hundreds of years (e.g., Phillips \& Hartmann \cite{phillips}, Baliunas et al. \cite{baliunas}, Saar \& Brandenburg \cite{saar1},  \cite{saar2}, Lockwood et al. \cite{lockwood}; we also refer to our own works, e.g., Strassmeier \cite{Klaus_rev} for a brief review of stellar cycle observations, Ol\'ah et al. \cite{olahetal} for a sample of 20 well-studied spotted stars with multiple and changing brightness cycles, and Vida et al. \cite{vida} for cycle evidence from \emph{Kepler} photometry). Not surprisingly, cycle related amplitudes and effective temperatures of such stars are, at best, difficult to obtain because of the long periods and small amplitudes. Even for the Sun, the light variation in the Johnson $V$ band over its 11 yr cycle does not amount to more than a few thousandths of a magnitude (Ol\'ah et al. \cite{mendoza}). In a recent investigation of the previous three solar cycles, Fr\"ohlich (\cite{frohlich}) showed that  the long-term solar variability requires a global solar temperature change, likely originating from the changing strength of its magnetic activity. The solar effective temperature appears lower during low activity states, and even differed by about 0.25\,K between the two minima in 1996 and 2008.

For other stars, cycles can show much higher amplitudes than the Sun, typically a few tenths of a magnitude in $V$. Such an amplitude range is still explainable with more or fewer (cool) spots on the stellar surface. The observed rotational modulation of active stars is indeed due to these apparent temperature changes caused by cool starspots moving in and out of sight with ample evidence from multicolor photometry in the literature (see, e.g., Strassmeier \cite{Klaus_rev} and the many references therein).  However, in some cases, the amplitude of the long-term variation is far larger than the rotational modulation itself and exceeds the limits from spot additivity (which is also subject of the stellar inclination). The question arises if it is possible at all to explain such changes with increasing or decreasing numbers of cool spots and/or warm faculae. As an example, a one-magnitude $V$-brightness variation of a late-type star, i.e., a flux change of 2.5 times, implies a change of several hundred degrees of its disk-integrated temperature. Three active giant stars exhibit such huge brightness changes and have also been observed  for decades with high cadence using multicolor photometry, which may allow the surface temperature determination. These stars are the topic of this paper: IL~Hya, XX~Tri, and DM\,UMa. Recently, Tang et al. (\cite{sumin}) discovered three more K giants with long-term 1 mag light variations from the Digital Access to a Sky Century at Harvard (DASCH) database, which could be RS~CVn-like stars. 

IL~Hya (HD~81410,  HIP~46159) is the K0III-IV star primary of a synchronized SB2-type RS~CVn binary system (P$_\mathrm{rot\approx orb}$=12.9 days), which has shown a 0\fm9 total brightness change in $V$ over the last 42 years. Its secondary star was first detected by Donati et al. (\cite{donati}) who gave a luminosity ratio of the components of about 1:23-24 and $5900\pm250$\,K for the temperature of the secondary star, making a difference of only 0\fm01 in $V-I_C$. Weber \& Strassmeier (\cite{michi_klaus}) derived the first SB2 orbital solution and published the first Doppler map of the primary star. We refer to these two papers for the stellar parameters of IL~Hya. The distance of the system is 105~pc (van Leeuwen \cite{hipparcos}). 
The most recent orbital parameters were given by Fekel et al. (\cite{fekeletal}). 

The brightness variations of IL~Hya studied by Ol\'ah et al. (\cite{olahetal}) using 20 years of continuous measurements yielded two spot cycles and a long-term trend. One cycle had a timescale of 4.4~years, and the other grew slowly from 2.7 to 3 years. Just recently, K\H{o}v\'ari et al. (\cite{kovarietal}) found solar-type surface differential rotation with $\Delta\Omega / \Omega$=0.05 shear parameter from a time-series Doppler imaging study of IL\,Hya, as well as indications of meridional circulation with poleward surface flow on the order of 100\,m\,s$^{-1}$.

XX~Tri (HD~12545, HIP~9630) is a K0III star in a synchronized SB1-type RS~CVn binary system ($P_\mathrm{rot\approx orb}$=24.0\,d), with a long-term total brightness change over 1 magnitude, in a distance of 160~pc (van Leeuwen \cite{hipparcos}). The mass function is small (0.0100$\pm0.0004$, Strassmeier, \cite{Klaus_xx}), the secondary star has not been detected in the spectra. The latest known orbital solution is given in Strassmeier (\cite{Klaus_xx}) where all previous information of the system is listed. A huge polar spot was derived from Doppler imaging, which is the largest spot ever found in an active star to date. In Strassmeier (\cite{Klaus_xx}) it is stressed that without warm spots the light curve amplitudes and the long-term brightness modulation could not be described. In the time of the huge polar spot in 1998, the star was its brightest state until that time,  but by 2009 it became more than 0.2 mag brighter. Using photometry from 21 consecutive years Ol\'ah et al. (\cite{olahetal}) found a long-term trend and only a weak, $\approx$3.8-year cycle in the second half of the dataset.

DM~UMa (BD +61$\degr$1211, HIP~53425) is also a single-lined RS~CVn binary (K0-1 IV-III, $P_\mathrm{rot\approx orb}$=7.49 days) at a distance of 139~pc (van Leeuwen \cite{hipparcos}).  The star has a 28-year photometric dataset collected from the literature together with a long series of new observations taken between 1988-2009, all presented by Rosario et al. (\cite{rosario_dmuma}). The secondary component is not detected, the mass function is low ($\approx$0.016, Crampton et al. \cite{crampton}). In this paper we do not use the Kavalur observations from Mohin et al. (\cite{mohin}), since they used a different comparison star than for all other observations, and the magnitude differences between the two comparison stars were not observed properly. A very valuable part of the Rosario et al. (\cite{rosario_dmuma}) data is the 17-year continuous series of parallel $UBVR_CI_C$ measurements. There is only one Doppler image for DM~UMa to date, from 1992-93; it was calculated by Hatzes (\cite{artie}) and shows a large polar spot plus three lower latitude features. In addition, the inclination of the system was also determined as $\approx$40\degr. Recently, Tas \& Evren (\cite{tas_evren}), based on the data from the literature supplemented with their own $UBVR$ data taken between 1997-2008, determined a $15.1\pm0.7$ yr cycle from the $V$ measurements and studied the time behavior of the spot minima. The star has a long-term variation with an amplitude of about 0\fm6 in $V$. 

In Sect.~\ref{obs}, we describe the new photometric data. In Sect.~\ref{results_ilhya}, Sect.~\ref{results_xxtri}, and Sect.~\ref{results_dmuma} the results are given in several subsections for the three stars. We show and discuss  the positions of the stars on the H-R diagram in Sect.~\ref{hrd_pos}, and in Sect.~\ref{disc} we present a summary and a broader discussion of the results.

\section{New observations, applied data bases, and calibrations}\label{obs}

\subsection{Long-term phase-resolved photometry}

The bulk of the photometry used in the analysis in this paper was taken with the 0.75m Vienna-Potsdam Automatic Photoelectric Telescope (APT) {\sl Amadeus} at Fairborn Observatory in southern Arizona (Strassmeier et al. \cite{apt0}) between 1993-2013. The APT observations through 1996 were published by Strassmeier et al. (\cite{aptdata}) together with a full list of data from the literature, which were added to our current dataset. The brightness information for IL~Hya,  XX~Tri, and DM~UMa now spans 42, 28, and 34 years, respectively. 

All observations of IL~Hya were taken in the $UBV(RI)_C$ bands, except Eggen's  (\cite{eggen}) $R$ and $I$ data which were  observed in the Kron-Eggen system and which we transformed to the Cousins system with the equations of Sandage~(\cite{sandage}). The same comparison star, \object{HD~81904}, was used by all observers. Its $V$ brightness and color indices are originally from Cameron (\cite{cameron}): $V$=8\fm03, $U-B$=0\fm72, $B-V$=0\fm98, $V-R_C$=0\fm52, and $V-I_C$=0\fm99; our APT results in $V(RI)_C$ are within 0\fm01 of these values. For the determination of effective temperatures at maximum and minimum brightness, i.e., minimum and maximum spot coverage, we selected 35 $V$ and $I_C$ light curves covering all 42 years of observations. Additional  $V,I_C$ observations at mid-brightness of the earliest light curves measured in the Cousins system on two nights, were also used to calculate temperatures. 

The observations of XX~Tri were taken in the $UBV(RI)_C$ colors using \object{HD~12478} as comparison star. Its $V$=7\fm78, $V-R_C$=0\fm72, and $V-I_C$=1\fm38 are determined from our APT data while $U-B$=1\fm46 and $B-V$=1\fm45 are from  Strassmeier \& Ol\'ah (\cite{strass_olah}). We selected 27 light curves  in $V$ and $I_C$ for effective temperature determination at maximum and minimum brightness. 

Observations of DM~UMa were mostly taken from the literature. In early 2014, we obtained 14 new $V,I_C$ measurements that were made six years after the last dataset was presented by Rosario et al. (\cite{rosario_dmuma}).  This left a six-year gap between 2008 and 2013. The APT solution allowed us to derive the brightness of its comparison star \object{HD~95362} to $V=9\fm05\pm0.03$ and $I_C=7\fm89\pm0.04$. 
For DM~UMa, we used 16 selected light curves in $V$ and $I_C$ from the literature and added the few most recent data points from the APT obtained in 2014.

\subsection{Auxiliary photometry}

We extracted all $JHK_s$ data from the Two Micron All-Sky Survey (2MASS; Skrutskie et al. \cite{2mass}), from the Wide-field Infrared Survey Explorer (WISE, Wright et al. \cite{wright}), and from the AKARI/IRC (Ishihara et al. \cite{akari}) observations of all three stars. These data were obtained between 1998--2010, and were used to check if there was any dust in the systems through spectral-energy distribution (SED) fits. During this time the maximum brightness of IL~Hya varied just by 0\fm14, and the rotational modulation was low, so not much variation is suspected in the mid- and near-infrared. The situation is the same for DM~UMa. The maximum brightness of XX~Tri varied by more than 0\fm3 with sometimes very large ($>$0\fm6) rotational modulation. The amplitudes of the observed light curves, that likely originate from cool spots, decrease towards the longer IR wavelengths, but the near- and mid-infrared data of XX~Tri are still very useful for checking for dust contribution.

\subsection{High-resolution spectroscopy}

Phase-resolved spectroscopic observations of IL\,Hya were obtained with the McMath-Pierce solar telescope during a single 
70-night observing run from November 1, 1996, through January 8, 1997.  An 800$\times$800-pixel 15$\mu$m TI CCD was used in conjunction with the Milton-Roy grating no. 1 to give an effective wavelength resolution of 0.15~\AA \ and a usable wavelength range of 55~\AA \ centered at 6435~\AA. Bias subtraction, flat-field division, wavelength calibration, and
continuum rectification were performed on the raw spectra with tIRAF (distributed by NOAO).  Thorium-argon comparison
spectra were obtained each night at intervals of one to two hours to ensure an accurate wavelength calibration. 

High resolution ($R$=55,000) spectroscopy obtained with the STELLA \'Echelle Spectrograph (SES) at the robotic 1.2m STELLA-I telescope in Tenerife (Strassmeier et al. \cite{stella}), is used to redetermine the metallicity for XX\,Tri. 

\subsection{Temperature calibration from $V-I_C$ data}\label{tempcal}

Throughout this paper we use the empirical temperature calibration of Worthey \& Lee (\cite{worthey}). The calibration is based on color-color relations from measured Johnson/Cousins photometry of 4496 stars with different spectral types, of which about 2090 stars have usable metallicity.  The results are given in tabular form. We interpolated the tables using our measured $V-I_C$ color indices and took into account the [Fe/H] and log~$g$ values of the stars. We estimate an error of about 50\,K in our resulting $T_\mathrm{eff}$ values originating from the error of the observations and from the interpolation. 

The studied objects XX~Tri and DM~UMa are SB1-type binaries without any measured information about the secondary stars, therefore we suppose that the light originating from these secondaries is negligible. In the case of IL~Hya the given properties of the secondary star (see  Section~\ref{intro}) indicates an about 0\fm01 change in the $V-I_C$ color index which translates to about a 20\,K difference in the resulting temperatures as a shift only, thus we do not take into account this small and uncertain correction.

%-------------------------------- Fig. 1.
\begin{figure}[t]
   \centering
   \includegraphics[width=8cm]{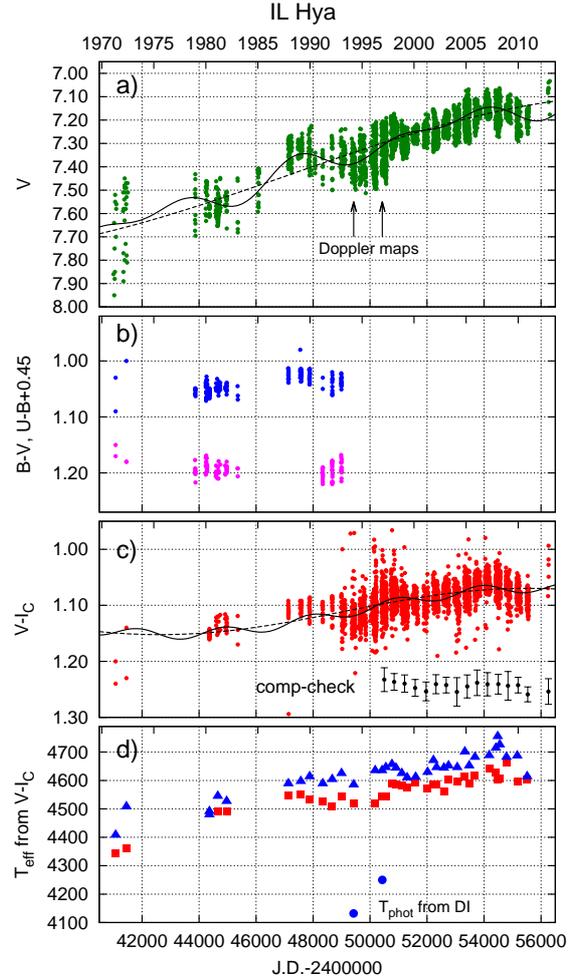}
   \caption{IL~Hya light, color, and temperature curves. {\bf a:} $V$ data (dots) and their fits with a  trend (dashed line; an artificial very long period fit), and with co-added modulations of timescales of 9~yr and 14~yr (line).The dates of the two Doppler images are indicated. {\bf b:} $B-V$ (upper dots) and $U-B$ (lower dots). The $U-B$ data were artificially shifted by +0\fm45 for easier presentation. {\bf c:} $V-I_C$ data (dots) and their fits with a trend (dashed line), and with co-added modulation of  a marginally significant variability on a timescale of $\approx$7~yr (line).  Seasonal average comp-check values are also plotted shifted by +1\fm20  for easier presentation. {\bf d:} Effective temperatures from $V-I_C$ at maxima (triangles) and minima (squares) of the light curves.} 
 \label{data_temp}
\end{figure}

\section{Results for IL~Hya}\label{results_ilhya}

\subsection{Cyclic brightness and color changes}

Figure~\ref{data_temp} shows the long-term behavior of the $V$ light and $V-I_C$ color index during the past 42 years. When subtracting the long-term $V$-band trend, a variation on a decade-long timescale (9--14~yr) with an amplitude of $\approx$0\fm1 becomes evident. The amplitude of the long-term change in the mean $V-I_C$ color is about six times smaller than in $V$ and about 0\fm10, which is similar to the highest amplitude of the rotational modulation ever seen in this color index. When subtracting its long-term trend, a variation of $\approx$7~yr with a $V-I_C$ amplitude of $\approx$0\fm025 remains. However, its period is far less significant than the period from $V$, but it is tempting to speculate that the 7 yr may be just half of the 9--14~yr from $V$. We note that the trends and periodicities in this paper were determined from multifrequency Fourier analysis (Koll\'ath \cite{mufran}). We emphasize that the cycle lengths referred to in this paper are in practice just timescales because there is evidence that cycle lengths can be variable in active stars (Ol\'ah et al. \cite{olahetal}). 

\subsection{Temperatures and luminosities from long-term photometry}\label{il_temp}

We chose eight representative multicolor light curves obtained within the shortest possible time interval, preferably during one stellar rotation in order to minimize the effect of an intrinsically changing spot coverage. These light curves cover the faintest and the brightest states of the star during the 42 years. Six light curves were chosen between the brightness extrema, given that observations were available in $UBV(RI)_C$ between 1971 and 1993. One light curve, extracted during maximum brightness in 2008, was taken only in $V$ and $I_C$.  $BV(RI)_C$  data from two consecutive nights (Cameron~\cite{cameron}) were averaged representing some middle (rotational) phase of the light curve between maximum and minimum. The maximum and minimum magnitudes of these light curves, and the two values at middle phases in four colors, are given in Table~\ref{maxmin_il}. 

\begin{table*}[t]
\caption{Maximum and minimum brightness and color indices of IL~Hya.}             
\label{maxmin_il}      
\centering                          
\begin{tabular}{l c c c c c c c c c }       
\hline\hline \noalign{\smallskip}               
 J.D. & $U$ & $B$ & $V$ & $R_C$ & $I_C$ & $U-B$ & $B-V$ & $V-R_C$ & $V-I_C$ \\     
        & max. min. & max. min. & max. min. & max. min. & max. min. & max. min. & max. min. & max. min. & max. min. \\ 
        &  [mag]      &  [mag]      & [mag]      &  [mag]      &  [mag]       &  [mag]      &  [mag]      &  [mag]      &  [mag]  \\   
                        \noalign{\smallskip}\hline \noalign{\smallskip} 
2441072 & 9.30\hspace{1.5mm}  9.66 & 8.60\hspace{1.5mm}  8.94 & 7.57\hspace{1.5mm}  7.85 & 6.94\hspace{1.5mm}  7.13 & 6.37\hspace{1.5mm}  6.61 & 0.70\hspace{1.5mm}  0.72 & 1.03\hspace{1.5mm}  1.09 & 0.63\hspace{1.5mm}  0.63 & 1.20\hspace{1.5mm}  1.24 \\
2441451 & 9.21\hspace{1.5mm}  9.56 & 8.48\hspace{1.5mm}  8.83 & 7.48\hspace{1.5mm}7.78 & 6.88\hspace{1.5mm}  7.10 & 6.34\hspace{1.5mm} 6.55 & 0.73\hspace{1.5mm}  0.73 & 1.00\hspace{1.5mm}  1.05 & 0.60\hspace{1.5mm}  0.68 & 1.14\hspace{1.5mm}  1.23 \\
2444363$^{\it a}$ & ---& 8.669 & 7.609 & 6.974 & 6.453 & --- & 1.060  & 0.635  & 1.156 \\
2444364$^{\it a}$ & ---& 8.630 & 7.575 & 6.974 & 6.425 & --- & 1.055  &  0.601 & 1.150 \\
2444660 & 9.32\hspace{1.5mm}  9.43 & 8.58\hspace{1.5mm}  8.67 & 7.54\hspace{1.5mm}  7.62 & 6.96\hspace{1.5mm}  7.02 & 6.42\hspace{1.5mm}  6.47 & 0.74\hspace{1.5mm}  0.76 & 1.04\hspace{1.5mm}  1.05 & 0.58\hspace{1.5mm}  0.60 & 1.12\hspace{1.5mm}  1.15 \\
2444970 & 9.30\hspace{1.5mm}  9.37 & 8.55\hspace{1.5mm}  8.64 & 7.51\hspace{1.5mm}  7.59 & 6.91\hspace{1.5mm}  6.99 & 6.38\hspace{1.5mm}  6.44 & 0.74\hspace{1.5mm}  0.74 & 1.04\hspace{1.5mm}  1.06 & 0.60\hspace{1.5mm}  0.60 & 1.13\hspace{1.5mm}  1.15 \\
2448339 & 9.10\hspace{1.5mm}  9.30 & 8.36\hspace{1.5mm}  8.53 & 7.33\hspace{1.5mm}  7.48 & 6.77\hspace{1.5mm}  6.90 & 6.23\hspace{1.5mm}  6.35 & 0.74\hspace{1.5mm}  0.77 & 1.03\hspace{1.5mm}  1.05 & 0.56\hspace{1.5mm}  0.58 & 1.10\hspace{1.5mm}  1.13 \\
2448672 & 9.03\hspace{1.5mm} 9.33 & 8.30\hspace{1.5mm}  8.56 & 7.28\hspace{1.5mm}  7.50 & 6.71\hspace{1.5mm}  6.91 & 6.19\hspace{1.5mm}  6.36 & 0.73\hspace{1.5mm}  0.77 & 1.02\hspace{1.5mm}  1.06 & 0.57\hspace{1.5mm}  0.59 & 1.09\hspace{1.5mm}  1.14 \\
2449000 & 9.00\hspace{1.5mm}  9.21 & 8.27\hspace{1.5mm}  8.47 & 7.24\hspace{1.5mm}  7.43 & 6.68\hspace{1.5mm}  6.85 & 6.16\hspace{1.5mm}  6.31 & 0.73\hspace{1.5mm}  0.74 & 1.03\hspace{1.5mm}  1.04 & 0.56\hspace{1.5mm}  0.58 & 1.08\hspace{1.5mm}  1.12 \\
2454480$^{\it b}$ &--- &---& 7.07\hspace{1.5mm}  7.22 & --- & 6.05\hspace{1.5mm}  6.13 & --- & --- &--- & 1.02\hspace{1.5mm}  1.09 \\
\noalign{\smallskip}\hline                  
\end{tabular}
\tablefoot{ $^{\it a}$~Average magnitudes on two consecutive nights from 10 and 17 data points, with a standard deviation of about 0\fm006. $^{\it b}$~Brightness at maximum for reference.} 
\end{table*}

Over 42 years the brightness of IL~Hya varied by about 0\fm9 in $V$ from the deepest minimum to the highest maximum (Fig.~\ref{data_temp}a). From  this it can be seen, that the $U-B$ and $B-V$ colors remained almost constant within 0\fm01-0\fm02 (Table~\ref{maxmin_il}), but the $U$ and $B$ data covered only a small fraction in time. The $V-R_C$ color index showed a small blueing of 0\fm05-0\fm06 until 1993 (after 1993 we have no $U$, $B$, and $R_C$ data), while the $V-I_C$ color index became continuously bluer by 0\fm15-0\fm18 during the brightening of the star (see Fig.~\ref{data_temp}b, c).  All color indices are redder at minimum brightness than at maximum brightness, verifying that the rotational modulation is caused by cool spots (Poe \& Eaton \cite{poe_eaton}). The observed maximum $V$ brightness, i.e., the least spotted or even unspotted state, varied altogether by 0\fm5 over the 42\,yr (Table~\ref{maxmin_il}).

We convert our $V-I_C$ indices to an effective temperature based on the calibration of Worthey \& Lee (\cite{worthey}). This was done for times of maximum and minimum brightness of all consecutive rotational modulations. Table~\ref{temp_lum_il} lists the derived temperatures, bolometric magnitudes, and luminosities for 37 epochs. Unfortunately, no data were available in  $U$ and $B$ after 1993. Figure~\ref{data_temp} displays the effective temperatures at maximum and minimum of the rotational modulation, parallel with the light and color index curves. We note that at a distance of 105\,pc, IL Hya is likely located inside the
low-density cavity of the Local Bubble (Reis et al. \cite{reis3_etal}) and, therefore, reddening was neglected. Based on the observational photometric error of 0\fm005 for $V,I_C$, we are confident that the relative temperatures are precise to within $\approx$50\,K (see Section~\ref{tempcal}). 

The temperatures appear systematically lower by about $50-150$\,K at minimum brightness of the light curves, i.e.,, at maximum spot visibility. Therefore, it appears that the temperature change follows the change in the spot visibility, being cooler when more spots are in view (Table~\ref{temp_lum_il}). This result is fully consistent with the observed temperature amplitudes from line-depth ratios, e.g., for the RS~CVn binaries VY~Ari, IM~Peg, and HK~Lac, which were measured to be 177\,K, 119\,K, and 127\,K, respectively, during a single stellar rotation (Catalano et al. \cite{catalano}). The long-term, 0\fm1, $V-I_C$ color change can be larger and corresponds to $150-200$\,K. It is due to a changing activity level, not just the number of spots, and it results in a globally different spot temperature. In this respect, the term ``spot'' refers to an active region likely composed of cool (spots) and hot (faculae) regions whose intensity ratio may change at different levels of activity (Bonomo \& Lanza \cite{bonomo_lanza}). Such a long-term spot temperature change of about 400\,K was already observed on IM~Peg from multicolor photometry by Rib\'arik et al. (\cite{ribarik}): spots became hotter with the overall brightening of the star indicating more faculae relative to spots. We suppose the same to be true on IL~Hya.

Because of the huge increase in brightness (0\fm9 in $V$) from absolute minimum to overall maximum, the corresponding effective temperature of the star varied by about 300$\pm$50\,K. If we consider just the observed maxima during the 42 years of observations, the brightness change in $V$ was about 0\fm5 and the corresponding temperature change not more than 200\,K, i.e., comparable to the highest amplitude ever seen from rotational modulation. As lowest maximum brightness we chose $V$=7\fm54 observed in 1981 (JD\,2,444,660, Table~\ref{maxmin_il}). We note that the maximum of the first observed light curve was fainter than this by 0\fm03, but the corresponding $I$ data were taken in the Kron-Eggen system and are more uncertain owing to the applied transformation (see Sect.\ref{obs}). The highest maximum brightness was $V$=7\fm07 in 2008 (JD\,2,454,480, Table~\ref{maxmin_il}). Using again the tables of Worthey \& Lee (\cite{worthey}) with $\log g$=3.0, [Fe/H]=$-$0.9, and the observed $V-I_C$ values, we calculated $m_{\mathrm{bol,max}}$ = 7\fm045 and 6\fm69 for 1981 and 2008, respectively. Thus, the bolometric flux between the lowest and highest light-curve maxima increased by 39\%\ within 27 years, and from this only $\approx$20\%\ can be explained by a temperature change of about 200\,K.  
        
\begin{table}
\caption{Temperatures and luminosities of IL~Hya from $V-I_C$ at maximum and minimum brightness, using $\log g$=3.0, [Fe/H]=--0.9.}             
\label{temp_lum_il}      
\centering                          
\begin{tabular}{l l l l l}       
\hline\hline\noalign{\smallskip}    
      J.D.  &  $T_\mathrm{eff, max}$ & $T_\mathrm{eff, min}$ & $m_{\mathrm{bol,max}}$ $L/L_{\odot, max}$ & $m_{\mathrm{bol,min}}$  $L/L_{\odot, min}$ \\    
               &   [K]  & [K]  & [mag] &  [mag]   \\
\noalign{\smallskip}\hline\noalign{\smallskip}
2441072  &  4408 & 4345 & 6.98\hspace{2mm} 14.0$\pm$1.5 & 7.22\hspace{2mm} 11.3$\pm$1.5 \\  
2441451  &  4509 & 4361 & 6.96\hspace{2mm} 14.3$\pm$1.5 & 7.16\hspace{2mm} 11.9$\pm$1.5 \\  
2444363$^{\it a}$ &  \multicolumn{2}{c}{4481}  &   \multicolumn{2}{c}{7.07\hspace{2mm} 12.9$\pm$1.4} \\ 
2444364$^{\it a}$ &  \multicolumn{2}{c}{4492}  &   \multicolumn{2}{c}{7.04\hspace{2mm} 13.2$\pm$1.4} \\  
2444660  &  4546 & 4492 & 7.04\hspace{2mm} 13.2$\pm$1.4 & 7.09\hspace{2mm} 12.7$\pm$1.4 \\
2444970  &  4526 & 4492 & 7.00\hspace{2mm} 13.8$\pm$1.5 & 7.06\hspace{2mm} 13.0$\pm$1.5 \\ 
2447136  &  4589 & 4548 & 6.83\hspace{2mm} 16.2$\pm$1.8 & 6.92\hspace{2mm} 14.8$\pm$1.8 \\
2447570  &  4597 & 4552 & 6.85\hspace{2mm} 15.8$\pm$1.8 & 6.88\hspace{2mm} 15.5$\pm$1.8 \\
2447880  &  4614 & 4532 & 6.80\hspace{2mm} 16.6$\pm$1.9 & 6.93\hspace{2mm} 14.7$\pm$1.9 \\
2448339  &  4585 & 4526 & 6.86\hspace{2mm} 15.7$\pm$1.7 & 6.97\hspace{2mm} 14.1$\pm$1.7 \\ 
2448672  &  4605 & 4509 & 6.82\hspace{2mm} 16.2$\pm$1.8 & 6.98\hspace{2mm} 14.0$\pm$1.8 \\
2449000  &  4625 & 4546 & 6.80\hspace{2mm} 16.6$\pm$1.8 & 6.93\hspace{2mm} 14.6$\pm$1.8 \\
2449425  &  4585 & 4518 & 6.82\hspace{2mm} 16.2$\pm$1.8 & 6.90\hspace{2mm} 15.1$\pm$1.8 \\
2450180  &  4635 & 4518 & 6.78\hspace{2mm} 16.9$\pm$1.9 & 6.97\hspace{2mm} 14.2$\pm$1.9 \\
2450420  &  4635 & 4542 & 6.79\hspace{2mm} 16.7$\pm$1.9 & 6.92\hspace{2mm} 14.9$\pm$1.9 \\
2450555  &  4646 & 4542 & 6.79\hspace{2mm} 16.8$\pm$1.9 & 6.94\hspace{2mm} 14.5$\pm$1.9 \\
2450770  &  4657 & 4589 & 6.77\hspace{2mm} 17.0$\pm$1.9 & 6.89\hspace{2mm} 15.3$\pm$1.9 \\
2450910  &  4644 & 4585 & 6.78\hspace{2mm} 16.9$\pm$1.9 & 6.83\hspace{2mm} 16.1$\pm$1.9 \\
2451115  &  4625 & 4583 & 6.79\hspace{2mm} 16.8$\pm$1.9 & 6.86\hspace{2mm} 15.7$\pm$1.9 \\
2451300  &  4610 & 4575 & 6.80\hspace{2mm} 16.5$\pm$1.9 & 6.85\hspace{2mm} 15.8$\pm$1.9 \\
2451600  &  4612 & 4593 & 6.78\hspace{2mm} 16.8$\pm$1.9 & 6.82\hspace{2mm} 16.2$\pm$1.9 \\
2452020  &  4629 & 4573 & 6.74\hspace{2mm} 17.5$\pm$2.0 & 6.84\hspace{2mm} 16.0$\pm$2.0 \\
2452225  &  4671 & 4587 & 6.74\hspace{2mm} 17.5$\pm$2.0 & 6.86\hspace{2mm} 15.7$\pm$2.0 \\
2452335  &  4646 & 4585 & 6.78\hspace{2mm} 16.8$\pm$1.9 & 6.85\hspace{2mm} 15.9$\pm$1.9 \\
2452590  &  4644 & 4561 & 6.77\hspace{2mm} 17.1$\pm$1.9 & 6.81\hspace{2mm} 16.5$\pm$1.9 \\
2452755  &  4652 & 4605 & 6.74\hspace{2mm} 17.5$\pm$2.0 & 6.81\hspace{2mm} 16.4$\pm$2.0 \\
2453060  &  4646 & 4597 & 6.74\hspace{2mm} 17.6$\pm$2.0 & 6.82\hspace{2mm} 16.3$\pm$2.0 \\
2453325  &  4701 & 4614 & 6.72\hspace{2mm} 17.9$\pm$2.0 & 6.80\hspace{2mm} 16.6$\pm$2.0 \\
2453480  &  4652 & 4589 & 6.69\hspace{2mm} 18.3$\pm$2.1 & 6.82\hspace{2mm} 16.2$\pm$2.1 \\
2453680  &  4682 & 4618 & 6.70\hspace{2mm} 18.1$\pm$2.0 & 6.79\hspace{2mm} 16.7$\pm$2.0 \\
2454180  &  4688 & 4642 & 6.71\hspace{2mm} 18.0$\pm$2.0 & 6.75\hspace{2mm} 17.3$\pm$2.0 \\
2454415  &  4714 & 4627 & 6.68\hspace{2mm} 18.4$\pm$2.1 & 6.79\hspace{2mm} 16.8$\pm$2.1 \\
2454480  &  4756 & 4605 & 6.69\hspace{2mm} 18.3$\pm$2.0 & 6.76\hspace{2mm} 17.1$\pm$2.0 \\  
2454550  &  4726 & 4608 & 6.69\hspace{2mm} 18.3$\pm$2.1 & 6.77\hspace{2mm} 17.1$\pm$2.1 \\
2454785  &  4682 & 4663 & 6.73\hspace{2mm} 17.7$\pm$2.0 & 6.75\hspace{2mm} 17.4$\pm$2.0 \\
2455165  &  4809 & 4595 & 6.71\hspace{2mm} 18.1$\pm$2.0 & 6.82\hspace{2mm} 16.3$\pm$2.0 \\
2455510  &  4614 & 4605 & 6.71\hspace{2mm} 18.0$\pm$2.0 & 6.77\hspace{2mm} 17.0$\pm$2.0 \\
   \noalign{\smallskip}\hline
\end{tabular}
\tablefoot{$^{\it a}$ From average magnitudes in two consecutive nights at mid-brightness of the rotational modulation.} 
\end{table}
           
\subsection{Average photospheric temperature of IL~Hya from Doppler images in 1994 and 1996}\label{section_di}

%--------------------------------------- Fig. 2.
   \begin{figure}[]
   \centering
   \includegraphics[width=5cm]{Olah_fig2a.eps}\hspace{2mm}\raisebox{5mm}{\includegraphics[width=3cm]{Olah_fig2b.eps}}\vspace{3mm}
   \includegraphics[width=8cm]{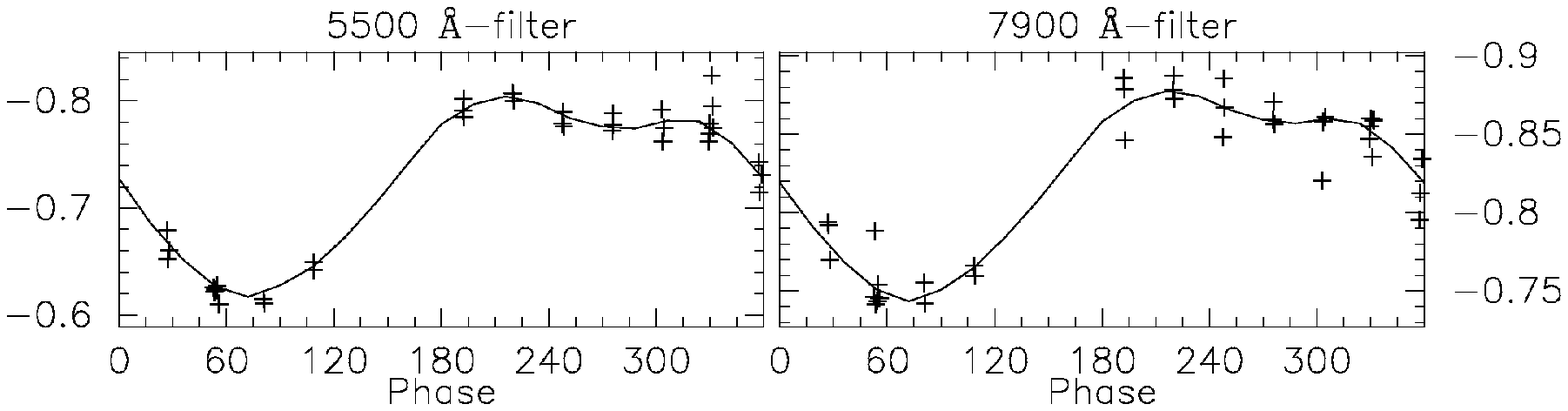}   
   \caption{Surface temperature distribution of IL~Hya from Doppler imaging in 1996. In this year the stars was at about its  medium brightness of the long-term change, marked in the upper panel of Fig.~\ref{data_temp}. \emph{Top left:} Mercator projection of the temperature map. \emph{Top right:} polar view of the temperature map. \emph{Bottom:} contemporaneous $V$ (5500\AA ) and $I_C$ (7900\AA ) light curves (plusses) and their fits (lines).}
         \label{doppler}
   \end{figure}

To obtain an independent value of the photospheric temperature, we use high-resolution spectroscopy and apply classical temperature Doppler imaging (e.g., Strassmeier \cite{Klaus_rev} and references therein). Observations were taken  with the 1.5m McMath-Pierce telescope on Kitt Peak in 1996-1997. A set of 12 spectra from 13 consecutive nights is available. This covers basically one stellar rotation  ($P_{\rm rot}\approx 13$\,d), which is the optimal time coverage for a  surface temperature map. The mean HJD of the data subset (covering the UT dates Dec 5--19, 1996) is 2,450,430.0. 
The wavelength range of the observed spectra includes two of the commonly used mapping lines, namely the Fe\,{\sc i} line at 6430\AA\ and the Ca\,{\sc i} line at 6439\AA, which have well-determined atomic parameters and are fairly free of telluric blends. Additionally, we used a total of 37 contemporaneous two-color photometric observations in $VI_C$, taken with Amadeus, one of the two 0.75m Vienna-Potsdam APTs at Fairborn Observatory in Arizona (Strassmeier et al. \cite{apt0}). These data are simultaneously inverted with the 2$\times$12 line profiles, but are given a weight of less than 10\%\ with respect to the line profiles.  The Doppler imaging code {\sc TempMap} (Rice et al. \cite{rice}) performs a full LTE spectrum synthesis by solving the equation of transfer through a set of Kurucz (\cite{kurucz_1993}) model atmospheres for a given set of chemical abundances and aspect angles. For more details see Rice \& Strassmeier (\cite{rice_strass_2000}) and Rice (\cite{rice2}). The stellar and atomic input parameters for Doppler imaging of IL~Hya were adopted from Weber \& Strassmeier (\cite{michi_klaus}). 
   
The resulting combined Fe\,{\sc i} and Ca\,{\sc i} Doppler image is shown in Fig.~\ref{doppler} along with the simultaneously fitted light curves. The map shows one small polar spot and other spots mostly between latitudes of $\approx$30\degr and $\approx$60\degr. The maximum temperature contrast is $\approx$800\,K below the unspotted photospheric surface. Our spot distribution map is very similar to the ones from 1994 and 1995 (see Weber \& Strassmeier \cite{michi_klaus}), except that in 1996 the lower latitude spots seem to be more dominant than the polar spot, in agreement with the increasing photometric amplitude. The average value of the recovered pixel temperatures over the unspotted part of the stellar surface is $\approx$4250\,K.  We also repeated the analysis of the spectra from 1994 from Weber \& Strassmeier (\cite{michi_klaus}) with the same parameters as for 1996 and obtained a photospheric temperature of 4130\,K. These temperatures are more than 300\,K cooler than those we calculated from the $V-I_C$ color indices, as plotted in Fig.~\ref{data_temp}d, and have an approximate uncertainty of 100-150\,K. While Doppler imaging yield only cool spots, the broadband colors used for temperature determination contain facular contribution, and this can be the reason for the discrepancy (see Section~\ref{disc} for more).

\subsection{Radius of IL~Hya}
 
Table~\ref{temp_lum_il} shows that the luminosity of IL~Hya, calculated from $m_{\mathrm{bol}}$ with $m_{\rm bol, \odot}=4.73$, increased proportionally to the brightening of the star, while the temperature change contributed only a moderate amount (200--250\,K) during the 42\,yr of observations.  Therefore, the luminosity increase of IL~Hya is only partly explained by the temperature change due to cool or warm spots. A more color-neutral luminosity increase would be expected in the case of stellar radius increase. Such radius changes at the same time could explain the nearly constant $B-V$ color. 

Using a different method we can independently determine the stellar radius from the measured rotational line broadening, $v\sin i$, by using values from the literature between 1981 to 1994 and with the inclination and rotational period from Weber \& Strassmeier (\cite{michi_klaus}). These values are summarized in Table~\ref{radius_il}, together with the relevant radii from  $T_{\rm eff}$ and $L$ from Table~\ref{temp_lum_il} at two common epochs. In 1981-82, the radius is found to be equal to within the error bars from the two methods, while in 1993-94 they are different, and possibly larger in 1993-94 than in 1981-82. Thus, a marginal change in the radius within 10 years might be claimed from the combination of $P_{\rm rot}$, $v\sin i$, and $i$ on the one hand, and the combination of $T_{\rm eff}$ and $L$ on the other hand, which are independent results.

\begin{table}
\caption{Stellar radius determinations of IL~Hya at different epochs and with different methods.}             
\label{radius_il}      
\centering                          
\begin{tabular}{cccccc}       
\hline\hline\noalign{\smallskip}    
      Years  &  $V$(max) & $v\,{\rm sin}\,i$ & $R_{\odot}$ & ref. & $R_{\odot}$ from  \\ 
                 & [mag] & [km\,$\rm s^{-1}$] & & & $T_{\rm eff}$ and $L$$^{\it b}$  \\
\noalign{\smallskip}\hline\noalign{\smallskip}
      1981-82   & 7.54 & $15\pm5$    & $4.6\pm1.6$$^{\it a}$ & 1 &  \\
                      & 7.54 & $22\pm2-3$ & $6.9\pm1$$^{\it a}$    & 2 & $5.9\pm0.3$ \\
      1987-90   & 7.26 & 22-23           & 6.9-7.2$^{\it a}$         & 3 &  \\                       
      1993-94   & 7.24 & $26.5\pm1$ & $8.1\pm^{0.9}_{0.7}$ & 4, 5 & $6.4\pm0.4$ \\
\noalign{\smallskip}\hline      
 \end{tabular}

\tablefoot{$^{\it a}$ We adopt $i=55\pm5$ deg and $P_{\rm rot}$=12.73\,d (Weber \& Strassmeier \cite{michi_klaus}). $^{\it b}$~This paper. References: (1) Collier \cite{andy}, (2) Fekel et al. \cite{fekel1}, (3) Pallavicini et al. \cite{pallavicini} and Randich et al. \cite{randich}, (4) Weber \& Strassmeier \cite{michi_klaus}, (5) Fekel et al. \cite{fekeletal}.} 
\end{table}

However, we must be aware that starspots affect the measured $v\sin i$ by distorting the disk-integrated line-profile shapes. This would be expected even when spots are symmetrically distributed on the surface so that they do not cause noticeable rotational modulation. IL~Hya has many spots including polar spotted regions (see Fig.~\ref{doppler}). In order to estimate the effect this would have on the $v\sin i$ measurement, we calculated the expected $v\sin i$ error with the parameters of IL~Hya assuming large polar spots as well as variously sized equatorial belts. During the calculations all the astrophysical parameters of IL~Hya ($v_{rot}$, inclination, etc.) were kept constant. This is detailed in Appendix~\ref{appA}. The results show that the $v\sin i$ change caused by a waxing or waning polar spot has a strong impact on the radius determination, while a spotted belt introduces almost no variation of the measured $v\sin i$ and thus has only a weak impact on the radius determination. It might have happened that, after 1981-82, when the star was continuously brightening, spots migrated up to the polar region increasing the percentage of the high latitude spots and pushing $v\sin i$ measurements to apparently higher values. As Fig.~\ref{vsini_test} shows, an increase in a polar cap of about 10\%\ can produce a $v\sin i$ increase of $\approx$1.5\,\kms. The measured $v\sin i$ of 26.5~\kms\ could thus be corrected to 25~\kms\ because of this polar spot effect, and this way the calculated radius in 1993-94 would be about 7.8\,$R_{\odot}$, which is still higher than our result. Therefore, the effect of a polar spot can only partly explain the similar (within 1$\sigma$) radii resulting from the two methods in 1981-82, and also only partly explain the deviating radius result in 1993-94, when the star was brighter. We note that at a time of similar brightness, i.e., in 1996, our Doppler map shows spots mostly at high latitudes. 

\section{Results for XX~Tri}\label{results_xxtri}

\subsection{Cyclic brightness and color changes}

\begin{figure}[]
   \centering
   \includegraphics[width=8cm]{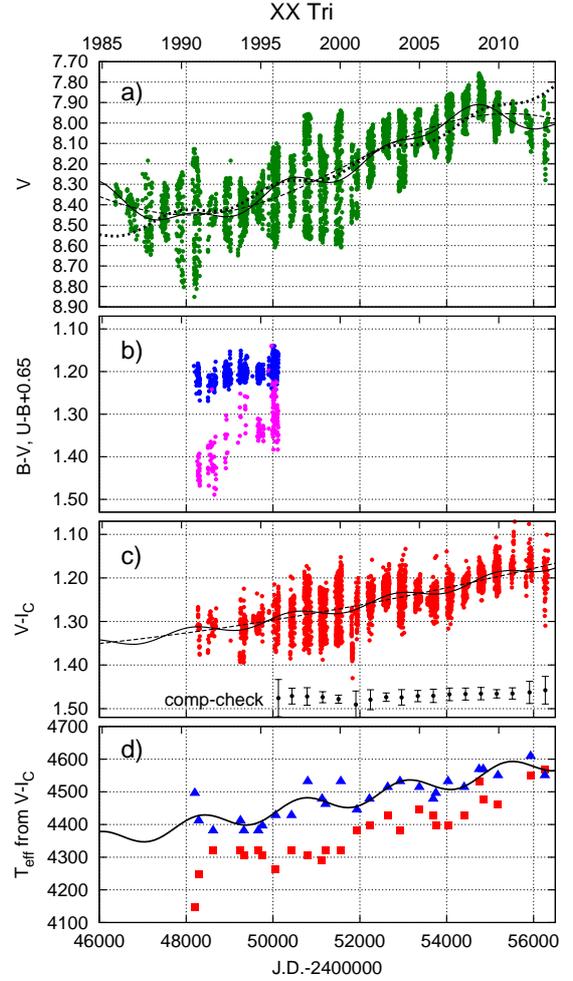}
   \caption{XX~Tri light, color, and temperature curves.  {\bf a:} $V$ data (dots) and their fits with a  trend (dashed line), with co-added modulations of timescales of 6~yr and 12~yr (line). Dotted line shows the cyclic term from the fit of the $V-I_C$ data. {\bf b:} $B-V$ (upper dots) and $U-B$ (lower dots). The $U-B$ data were artificially shifted by +0\fm65 for easier presentation. {\bf c:} $V-I_C$ data (dots) and their fits with a trend (dashed line), and with co-added modulation on a timescale of $\approx$6.5~yr (line).  The seasonal average comp-check values are plotted (shifted by +1\fm64 for presentation purpose) measured in the same time, and {\bf d:} effective temperatures from $V-I_C$ at maxima (triangles, overplotted with the 6.5 yr cycle originating from the $V-I_C$ fit) and minima (squares) of the light curves. See text for details.}
         \label{xx_data_temp}
   \end{figure}

XX~Tri has a long-term $V$-band modulation over 1\fm0 from the deepest minimum to the overall maximum. The rotational modulation can also be very large, up about 0\fm65 in $V$. The maximum light, i.e., the least spotted faces of the star, varied  by about half of the total change, 0\fm5. This is suggestive of a high inclination of the stellar rotation axis together with uneven spot distributions (see Fig.~\ref{xx_data_temp}). XX~Tri reached its maximum brightness in 2009 and during the next three years it became about 0\fm2 fainter. However, the $V-I_C$ color index reached its maximum, i.e., bluest face of the star, two years later in 2011 and only slightly reddened thereafter. The timescale of the long-term brightness change (called trend) is comparable to the length of our dataset. When removing this trend from the data, the residuals show a cycle of about $\approx$6 years and its integer multiple (Fig.~\ref{xx_data_temp}). The $V-I_C$ color index shows a comparable cycle of about 6.5 yr, again after removing its long-term trend.

%----------------------------------------- Fig. 3.    

\subsection{Temperatures and luminosities from long-term photometry}

%----------------------------------------- Fig. 4.       
   \begin{figure}[]
   \centering
   \includegraphics[width=8cm]{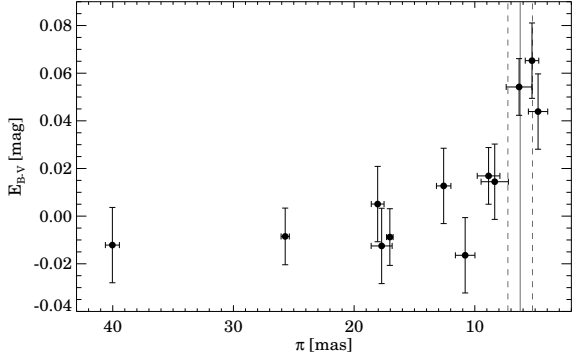}
   \caption{$E(B-V)$ reddening values as a function of trigonometric parallaxes taken from the \emph{Hipparcos}  catalog. The solid and the dashed vertical lines show the measured parallax for XX~Tri and its uncertainty, respectively.  The uncertainties of $E(B-V)$ are based on the values given by Vergely et al. (\cite{vergely}).}
         \label{xx_ebv}
   \end{figure}
   
From the entire dataset, we could extract only four sets of light curves with data in all five bands $UBV(RI)_C$. These light curves are in the range 1991-95. Their brightness and color indices at maximum and minimum light are given in Table~\ref{maxmin_xx}. During these five years the $U-B$ color index became bluer by about 0\fm15, while the other color indices also became bluer but not nearly as dramatic as $U-B$. $V$ and $I_C$ magnitudes at the times of the overall brightness maximum in 2009 are also added to Table~\ref{maxmin_xx} for further reference.

\begin{table*}[th]
\caption{Maximum and minimum brightness and color indices of XX~Tri from multicolor observations.}             
\label{maxmin_xx}      
\centering                          
\begin{tabular}{l c c c c c c c c c }       
\hline\hline \noalign{\smallskip}               
 J.D. & $U$ & $B$ & $V$ & $R_C$ & $I_C$ & $U-B$ & $B-V$ & $V-R_C$ & $V-I_C$ \\     
        & max. min. & max. min. & max. min. & max. min. & max. min. & max. min. & max. min. & max. min. & max. min. \\   
        &  [mag]      &  [mag]      & [mag]      &  [mag]      &  [mag]       &  [mag]      &  [mag]      &  [mag]      &  [mag]  \\       
       \noalign{\smallskip} \hline  \noalign{\smallskip}
2448285 & 	10.18\hspace{1.5mm} 10.75 &	9.41\hspace{1.5mm} 9.96& 8.23\hspace{1.5mm} 8.69& 7.59\hspace{1.5mm} 8.01& 6.96\hspace{1.5mm} 7.32& 0.77\hspace{1.5mm} 0.79& 1.18\hspace{1.5mm} 1.27&	0.64\hspace{1.5mm} 0.68& 1.27\hspace{1.5mm} 1.37\\
2448600 &	10.29\hspace{1.5mm} 10.53 &	9.59\hspace{1.5mm} 9.73& 8.36\hspace{1.5mm} 8.48& 7.72\hspace{1.5mm} 7.82&	7.07\hspace{1.5mm} 7.16& 0.70\hspace{1.5mm} 0.80& 1.23\hspace{1.5mm} 1.25&	0.64\hspace{1.5mm} 0.66& 1.29\hspace{1.5mm} 1.32\\
2449670 &	10.18\hspace{1.5mm} 10.40 &	9.51\hspace{1.5mm} 9.68& 8.32\hspace{1.5mm} 8.47& 7.68\hspace{1.5mm} 7.82&	7.04\hspace{1.5mm} 7.15& 0.67\hspace{1.5mm} 0.72& 1.19\hspace{1.5mm} 1.21&	0.64\hspace{1.5mm} 0.65& 1.28\hspace{1.5mm} 1.32\\
2450050 &	9.91\hspace{1.5mm} 10.46 &	9.29\hspace{1.5mm} 9.80& 8.13\hspace{1.5mm} 8.60& 7.50\hspace{1.5mm} 7.92&	6.87\hspace{1.5mm} 7.22& 0.62\hspace{1.5mm} 0.66& 1.16\hspace{1.5mm} 1.20&	0.63\hspace{1.5mm} 0.68& 1.26\hspace{1.5mm} 1.38\\
2454745$^{\it a}$ & --- &---&  7.76\hspace{1.5mm}  7.97& --- & 6.58\hspace{1.5mm}  6.77&--- & --- &--- & 1.18\hspace{1.5mm}  1.20\\
\noalign{\smallskip}\hline                  
\end{tabular}
\tablefoot{ $^{\it a}$ Brightness at maximum for reference.} 
\end{table*}

\begin{table}
\caption{Temperatures and luminosities of XX~Tri from $V-I_C$ at maximum and minimum brightness, using $\log g$=2.5, [Fe/H]=--0.27.}            
\label{temp_lum_xx}      
\centering                          
\begin{tabular}{l l l l l}       
\hline\hline \noalign{\smallskip}   
      J.D.  &  $T_\mathrm{eff, max}$ & $T_\mathrm{eff, min}$ & $m_{\mathrm{bol,max}}$ $L/L_{\odot}$ & $m_{\mathrm{bol,min}}$  $L/L_{\odot, min}$ \\  
                     &   [K]  & [K]  & [mag] &  [mag]   \\    
\noalign{\smallskip}\hline\noalign{\smallskip}
2448200& 4496& 4148& 7.64\hspace{2mm}  20.3$\pm$6.7& 8.01\hspace{2mm}  14.4$\pm$6.7 \\
2448290& 4412& 4247& 7.64\hspace{2mm}  20.3$\pm$6.7& 7.99\hspace{2mm}  14.7$\pm$6.7 \\
2448620& 4381& 4320& 7.75\hspace{2mm}  18.4$\pm$6.1& 7.82\hspace{2mm}  17.2$\pm$6.1 \\
2449250& 4412& 4320& 7.69\hspace{2mm}  19.4$\pm$6.4& 7.91\hspace{2mm}  15.8$\pm$6.4 \\
2449330& 4381& 4306& 7.74\hspace{2mm}  18.6$\pm$6.1& 7.86\hspace{2mm}  16.6$\pm$6.1 \\
2449660& 4381& 4320& 7.72\hspace{2mm}  18.9$\pm$6.2& 7.82\hspace{2mm}  17.2$\pm$6.2 \\
2449755& 4396& 4306& 7.65\hspace{2mm}  20.2$\pm$6.6& 7.84\hspace{2mm}  16.9$\pm$6.6 \\
2450060& 4428& 4262& 7.55\hspace{2mm}  22.0$\pm$7.3& 7.89\hspace{2mm}  16.1$\pm$7.3 \\
2450425& 4428& 4320& 7.60\hspace{2mm}  21.0$\pm$6.9& 7.81\hspace{2mm}  17.4$\pm$6.9 \\
2450800& 4532& 4306& 7.46\hspace{2mm}  23.9$\pm$7.9& 7.88\hspace{2mm}  16.3$\pm$7.9 \\
2451130& 4478& 4291& 7.58\hspace{2mm}  21.5$\pm$7.1& 7.89\hspace{2mm}  16.2$\pm$7.1 \\
2451210& 4462& 4320& 7.62\hspace{2mm}  20.8$\pm$6.8& 7.86\hspace{2mm}  16.6$\pm$6.8 \\
2451560& 4532& 4320& 7.45\hspace{2mm}  24.1$\pm$7.9& 7.93\hspace{2mm}  15.5$\pm$7.9 \\
2451931& 4445& 4381& 7.59\hspace{2mm}  21.2$\pm$7.0& 7.84\hspace{2mm}  16.9$\pm$7.0 \\
2452220& 4478& 4396& 7.50\hspace{2mm}  23.1$\pm$7.6& 7.71\hspace{2mm}  19.1$\pm$7.6 \\
2452640& 4514& 4428& 7.46\hspace{2mm}  23.9$\pm$7.9& 7.57\hspace{2mm}  21.6$\pm$7.9 \\
2452925& 4532& 4381& 7.44\hspace{2mm}  24.3$\pm$8.0& 7.71\hspace{2mm}  19.1$\pm$8.0 \\
2453370& 4514& 4445& 7.42\hspace{2mm}  24.8$\pm$8.2& 7.53\hspace{2mm}  22.4$\pm$8.2 \\
2453690& 4478& 4428& 7.44\hspace{2mm}  24.5$\pm$8.1& 7.53\hspace{2mm}  22.4$\pm$8.1 \\
2453750& 4496& 4396& 7.40\hspace{2mm}  25.3$\pm$8.3& 7.56\hspace{2mm}  21.9$\pm$8.3 \\
2454030& 4532& 4396& 7.36\hspace{2mm}  26.2$\pm$8.6& 7.57\hspace{2mm}  21.7$\pm$8.6 \\
2454400& 4514& 4428& 7.35\hspace{2mm}  26.5$\pm$8.7& 7.47\hspace{2mm}  23.7$\pm$8.7 \\
2454745& 4569& 4532& 7.28\hspace{2mm}  28.3$\pm$9.3& 7.46\hspace{2mm}  23.9$\pm$9.3 \\
2454840& 4569& 4478& 7.32\hspace{2mm}  27.3$\pm$9.0& 7.44\hspace{2mm}  24.5$\pm$9.0 \\
2455180& 4550& 4462& 7.37\hspace{2mm}  26.1$\pm$8.6& 7.50\hspace{2mm}  23.2$\pm$8.6 \\
2455930& 4609& 4550& 7.54\hspace{2mm}  22.2$\pm$7.3& 7.64\hspace{2mm}  20.4$\pm$7.3 \\
2456260& 4550& 4569& 7.51\hspace{2mm}  23.0$\pm$7.6& 7.77\hspace{2mm}  18.0$\pm$7.6 \\
\noalign{\smallskip}\hline
\end{tabular}
\end{table}

XX Tri is located at a distance of 160\,pc, well beyond the Local Bubble, thus its reddening cannot be neglected. In order to estimate the reddening of XX~Tri, we searched for stars in the vicinity of our target (with separations $<$3$^{\circ}$) with \emph{i}) Str\"omgren color indices including H$_\beta$ in the catalog compiled by Hauck \& Mermilliod (\cite{hauck}); and
\emph{ii}) trigonometric parallaxes measured by \emph{Hipparcos} (van Leeuwen \cite{hipparcos}). Based on the collected Str\"omgren data, we derived $E(b-y)$ color excesses for these objects by applying the appropriate calibration processes (Crawford \cite{craw1}, Crawford  \cite{craw2}, Olsen \cite{olsen}). Then, $E(B-V)$ values were computed from $E(b-y)$ via $E(B-V) = E(b-y) / 0.74$. Figure~\ref{xx_ebv} shows the color excesses as a function of trigonometric parallax. It  suggests $E(B-V)$ of $\sim$0\fm05 at the distance of XX~Tri. We note that when using the extinction map of Schlafly \& Finkbeiner (\cite{schlafly}) we obtain $E(B-V)$= 0\fm074 as an estimate for the total reddening in the direction of our target.

The $V-I_C$ color index-temperature calibration of Worthey \& Lee (\cite{worthey}) was again used to estimate the temperatures at maximum and minimum brightness of the rotational modulation at 27 epochs, whenever contemporaneous $V$ and $I_C$ measurements were available. The derived temperatures, bolometric magnitudes, and luminosities for the selected light curves are given in Table~\ref{temp_lum_xx}. The observational errors for this star are the same as for IL~Hya and thus the temperatures are also precise to within about 50\,K (see Section~\ref{tempcal}). Figure~\ref{xx_data_temp} shows the effective temperatures at maximum and minimum brightness of the modulated light curves, parallel with the light and color index curves. The fit of the $V-I_C$ color index (Fig.~\ref{xx_data_temp}c) with two components shows a long-term trend and a periodicity of about 6.5 years. This fit is also overplotted in panel~d in Fig.~\ref{xx_data_temp} and shows good agreement with the effective temperatures (blue triangles). However, the $V$ data is not well represented with this fit, indicating that cool spots are not the only causes of the long-term light variation. We note again that the brightness maximum occurred in 2009 while $V-I_C$ was  bluest in 2011-12.  The calculated temperature difference between maximum and minimum brightness is about $50-200$\,K, being cooler when more spots are in view at brightness minimum. This is again consistent with the independent temperature {\it amplitudes} observed during rotational modulation from line-depth ratios of line pairs with different temperature sensitivity of active RS~CVn giants (Catalano et al. \cite{catalano}, see also in Sect.~\ref{il_temp}). 

From absolute minimum to overall maximum the star becomes brighter in $V$ by more than one magnitude, and the  temperature change is about 450$\pm$50\,K. From the 28 years of available observations, we find that the brightness change from the maxima of the rotational modulation is, as for IL~Hya, about 0\fm5. The corresponding temperature change is $\approx$200\,K, i.e. twice as much as the average 100\,K due to the rotational modulation by cool spots. The faintest maximum brightness is $V$=8\fm36 observed in 1992 and the brightest maximum brightness is  $V$=7\fm76 in 2009 (Table~\ref{maxmin_xx}). Using the tables of Worthey \& Lee (\cite{worthey}) with $\log g$=2.5, [Fe/H]=$-0.27$, and our observed $V-I_C$ values, we calculate the bolometric magnitudes $m_{\rm bol}$=7\fm75 and 7\fm28 at these epochs, respectively. The bolometric flux between the faintest (JD\,2,448,600) and the brightest (JD\,2,454,745) maxima shows a 54\%\ increase, and from this only 16\%\ can be explained by temperature change. The brightest maximum is not the bluest one, and this makes a difference between the flux change; if we take $m_{\mathrm{bol,max}}$ when the star is the reddest (JD\,2,448,600) and the bluest (JD\,2,455,930), we obtain approximately an 20\%\ flux change both from $T_\mathrm{eff, max}$ and $m_{\mathrm{bol,max}}$ showing that this part of the flux variation is of temperature origin. The cause of the remaining 34\%\ of the flux increase is not clear. In Fig.~\ref{xx_data_temp}c, we plotted the seasonal average comparison-check $V-I_C$ color indices in order to demonstrate the constancy of the comparison star in both colors.  Strassmeier (\cite{Klaus_xx}) found from simultaneous photometry and spectroscopy obtained in 1997-98, that the light and color index curves could not be fitted together with the Doppler image if only cool spots were allowed; about 3.5\%\ of the total stellar surface had to be covered with regions about 350\,K hotter than the photosphere. After 1997-98 the star brightened by about 0\fm2 in $V$ to its all-time maximum brightness in 2009.
  
\section{Results for DM~UMa}\label{results_dmuma}

\subsection{Cyclic brightness and color changes}

We show all published photometry and our new $V$ and $V-I_C$ observations in Fig.~\ref{dm_data_temp}. DM~UMa showed nearly constant mean $B-V$ during the 28 years of observation (but still rotationally modulated) while its $V$ magnitude changed more than 0\fm6 in total (Rosario et al.  \cite{rosario_dmuma}). The low inclination of about $40\degr$ or even lower (Hatzes, \cite{artie}), suggests spottedness at high latitudes. Therefore, the amplitudes of the rotational modulation remained relatively small (0\fm1--0\fm2). After removing the long-term trend from the $V$ data (plotted with dashed line in Fig.~\ref{dm_data_temp}a1) a cycle-like variation on a timescale of 17 years remained (Fig.~\ref{dm_data_temp}a2). Overplotting this modulation to all color indices (Fig.~\ref{dm_data_temp}b,c) we see a good match with the data. Tas \& Evren (\cite{tas_evren}) reported a $15.1\pm0.7$ years long cycle. No trace of long-term trend is seen in $B-V$, which spans the full length of the dataset except our new observations.

In the course of the 17 yr timescale variation the $U-B$ color index shows amplitudes of more than 0\fm05, indicating a strong variation within the facular/plage regions (area and/or temperature). $B-V$ and $V-I_C$ have much smaller amplitudes (about 0\fm02), but all color indices vary in phase. The amplitude of the rotational modulation is the highest in $U-B$ and smaller in $B-V$ and $V-I_C$. This indicates that the 17 yr as well as the rotational variation of DM~UMa is caused by temperature changes and both have strong contributions from hot surface regions. 

\subsection{Temperatures and luminosities from long-term photometry}   

DM~UMa is located fairly far from the Galactic plane ($b\sim$51$^{\circ}$) at a distance of $\sim$140~pc. 
Based on the dust map of Schlafly \& Finkbeiner (\cite{schlafly}) the total reddening along DM~UMa's line of sight is only $\sim$0\fm01. Therefore, we assume that the extinction is negligible.  
  
\begin{figure}[]
   \centering
   \includegraphics[width=8cm]{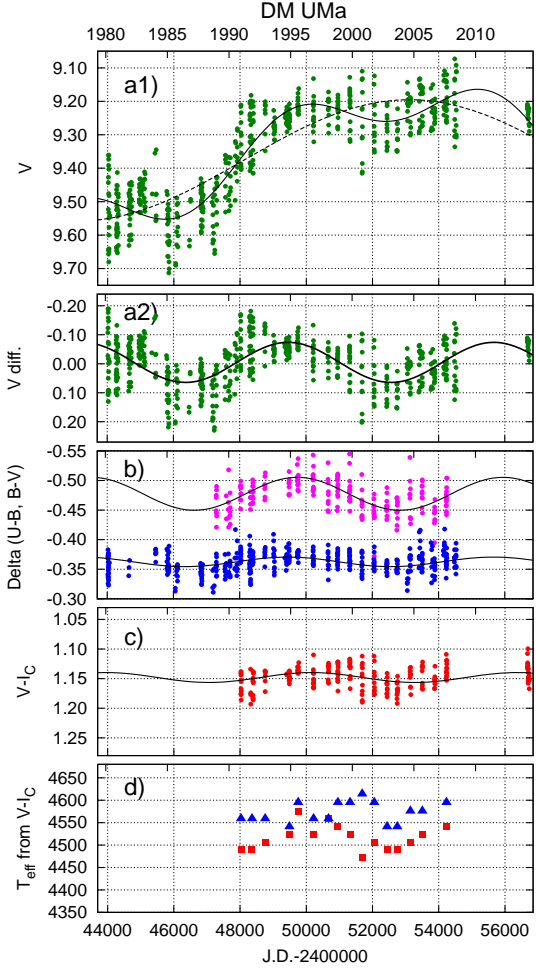}
   \caption{\emph{From top to bottom:} {\bf a1:} $V$ light curve with a fit of a long-term trend (dashed line), trend plus a 17 yr variation (full line), {\bf a2:} $V$ data difference from the long-term trend, showing the 17 yr variation from the fit, {\bf b:} $B-V, U-B$ color indices overplotted with the 17 yr variation, {\bf c:} $V-I_C$ color index overplotted with the 17 yr variation and {\bf d:} effective temperatures of DM~UMa between 1992-2007. See text for details.}
         \label{dm_data_temp}
   \end{figure}

As for the other two stars, we used the $V-I_C$ color index-temperature calibration of Worthey \& Lee (\cite{worthey}) to estimate the temperatures at maximum and minimum brightness of the rotational modulation. Sixteen suitable subsets of $V$ and $I_C$ light curves from 15 years between 1992-2007 are available. The derived temperatures, given in Table~\ref{temp_lum_dm}, are precise to within about 50\,K (see also Section~\ref{tempcal}) on the error estimates of $V-I_C$ given by Rosario et al. (\cite{rosario_dmuma}). Figure~\ref{dm_data_temp} shows the effective temperatures at maximum and minimum brightness of the modulated light curves, parallel with the light and color index curves. 
The average temperature of DM~UMa at maximum and minimum (blue triangles and red squares in Fig.~\ref{dm_data_temp}d, respectively) differ from each other by about 50\,K for most of the times, except when $U-B$ shows its maxima between 1994-1997. Then, the difference is smaller, a result of the low amplitude in $V-I_C$.  Although the uncertainty of the calculated temperature is on the same order as the variability itself,  the temperatures are systematically cooler at light curve minima. The facular contribution of the active regions with higher temperature relative to the photosphere varied with higher amplitude than the cooler spotted regions as seen from the temperatures at light curve minima, i.e., at maximum spot visibility. 

Unfortunately, contemporaneous $V$ and $I_C$ observations exist only for the $\sim$15 years when the overall brightness of the star did not change much. The rapid brightening of about 0\fm3-0\fm4 happened during a decade between 1980-1991, and the first useful $V$ and $I_C$ data for the temperature calculation were obtained in 1992. Our new observations from 2014 (six years later than the previously published data) show low-amplitude variability with considerable scatter which prevents us from applying the color index-temperature calibration. Because of the small temperature variation during the years with available $V-I_C$, we cannot estimate the contribution of the changes of temperature origin in the overall light variation. However, the $B-V$ color index, which also reflects the temperature variability as seen from the rotationally modulated light and color index curves in Rosario et al. (\cite{rosario_dmuma}), does not show the long-term change evident in $V$ color. This is at least suggestive that some other mechanism is responsible for that variation (see Fig.~\ref{dm_data_temp}).

%--------------------------------------  Fig. 5.

\begin{table}
\caption{Temperatures and luminosities of DM~UMa from $V-I_C$ at maximum and minimum brightness, $\log g$=3.0, [Fe/H]=0.0.}             
\label{temp_lum_dm}      
\centering                          
\begin{tabular}{l l l l l}       
\hline\hline \noalign{\smallskip}   
      J.D.  &  $T_\mathrm{eff, max}$ & $T_\mathrm{eff, min}$ & $m_{\mathrm{bol,max}}$ $L/L_{\odot}$ & $m_{\mathrm{bol,min}}$  $L/L_{\odot, min}$ \\  
              &   [K]  & [K]  & [mag] &  [mag]   \\  
\noalign{\smallskip}\hline\noalign{\smallskip}
2448030 &    4559 & 4489 & 8.71\hspace{2mm}  5.0$\pm$1.8  &   8.85\hspace{2mm}  4.3$\pm$1.8\\
2448360 &    4559 & 4489 & 8.70\hspace{2mm}  5.0$\pm$1.8  &   8.85\hspace{2mm}  4.3$\pm$1.8\\
2448760 &    4559 & 4506 & 8.70\hspace{2mm}  5.0$\pm$1.8  &   8.81\hspace{2mm}  4.5$\pm$1.8\\
2449490 &    4541 & 4524 & 8.73\hspace{2mm}  4.9$\pm$1.7  &   8.76\hspace{2mm}  4.7$\pm$1.7\\
2449755 &    4595 & 4576 & 8.69\hspace{2mm}  5.0$\pm$1.8  &   8.74\hspace{2mm}  4.8$\pm$1.8\\
2450220 &    4559 & 4524 & 8.69\hspace{2mm}  5.1$\pm$1.8  &   8.79\hspace{2mm}  4.6$\pm$1.8\\
2450670 &    4559 & 4559 & 8.68\hspace{2mm}  5.1$\pm$1.8  &   8.79\hspace{2mm}  4.6$\pm$1.8\\
2450955 &    4595 & 4541 & 8.70\hspace{2mm}  5.0$\pm$1.8  &   8.77\hspace{2mm}  4.7$\pm$1.8\\
2451320 &    4595 & 4524 & 8.65\hspace{2mm}  5.2$\pm$1.9  &   8.80\hspace{2mm}  4.6$\pm$1.9\\
2451690 &    4614 & 4473 & 8.64\hspace{2mm}  5.3$\pm$1.9  &   8.86\hspace{2mm}  4.3$\pm$1.9\\
2452055 &    4595 & 4506 & 8.64\hspace{2mm}  5.3$\pm$1.9  &   8.82\hspace{2mm}  4.5$\pm$1.9\\
2452440 &    4541 & 4489 & 8.70\hspace{2mm}  5.0$\pm$1.8  &   8.80\hspace{2mm}  4.5$\pm$1.8\\
2452750 &    4541 & 4489 & 8.72\hspace{2mm}  4.9$\pm$1.8  &   8.80\hspace{2mm}  4.5$\pm$1.8\\
2453145 &    4576 & 4506 & 8.70\hspace{2mm}  5.0$\pm$1.8  &   8.77\hspace{2mm}  4.7$\pm$1.8\\
2453505 &    4576 & 4524 & 8.67\hspace{2mm}  5.1$\pm$1.8  &   8.75\hspace{2mm}  4.8$\pm$1.8\\
2454235 &    4595 & 4541 & 8.64\hspace{2mm}  5.3$\pm$1.9  &   8.75\hspace{2mm}  4.8$\pm$1.9\\
\noalign{\smallskip}\hline
\end{tabular}
\end{table}
   
\section{Mass and age from position in the H-R diagram}\label{hrd_pos}

Magnitude-range brightness variations will likely also have an impact on a star's position in the H-R diagram (HRD). Its consequence could be systematically wrong masses and ages for such stars, in particular if they are single. The three targets of our present study should be easy to verify examples because they are binaries with known orbital elements and relatively solid estimations of the inclination of the rotational axis. A comparison with theoretical evolutionary tracks may thus quantify the impact of the presumably magnetically-induced brightness variations. Such a comparison is still a nearly virgin territory in stellar astrophysics, first of all, because low-mass tracks are usually computed without rotation or a magnetic field (see Spada et al. \cite{spada}), and, also because only just a handful of overactive stars have decade-long, continuous phase-resolved observations. We note that overactive stars, giants, and binaries were intentionally removed from the Mt Wilson H\&K survey (Wilson \cite{wilson}).  

Before we compare positions in the H-R diagram, we must re-evaluate the observed metallicities of our targets because these affect the computed evolutionary tracks significantly. Low metallicity for IL~Hya is understood from the low Ca and Fe abundance of 0.9~dex below solar derived from line-profile synthesis that is part of the Doppler-imaging process with {\sc TempMap} (Weber \& Strassmeier \cite{michi_klaus}). Randich et al. (\cite{randich}) gave an overall value of [Fe/H]=$-$0.5 from spectrum synthesis analysis fitting the observed iron lines near 6708\AA, which also supports the low metal content. The metallicity of XX~Tri was measured to be $-0.27\pm0.03$ from STELLA spectra and the PARSES synthesis routines (Jovanovic et al. \cite{parses}), i.e., it is also significantly metal-deficient. 

The question of the metal content in the photospheres of stars with strong magnetic activity is still open. Three-dimensional dynamic model atmospheres are likely needed to fully understand what we measure (e.g., Collet et al. \cite{collet}). Moreover, the line emission from plages may fill up the core of photospheric absorption lines and result in artificially low abundances of the elements on the surface. Morel et al. (\cite{morel1}) discussed the problem of the abundances of active stars and presented a comparison of different [Fe/H] results for five active stars, and of many elements for Arcturus obtained by different authors.

%----------------------------------- Fig. 6.
   \begin{figure}[]
   \centering
   \includegraphics[width=6.9cm]{Olah_fig6a.eps}
   \includegraphics[width=6.9cm]{Olah_fig6b.eps}
   \includegraphics[width=6.9cm]{Olah_fig6c.eps}
   \hspace*{-2mm} \includegraphics[width=7.35cm]{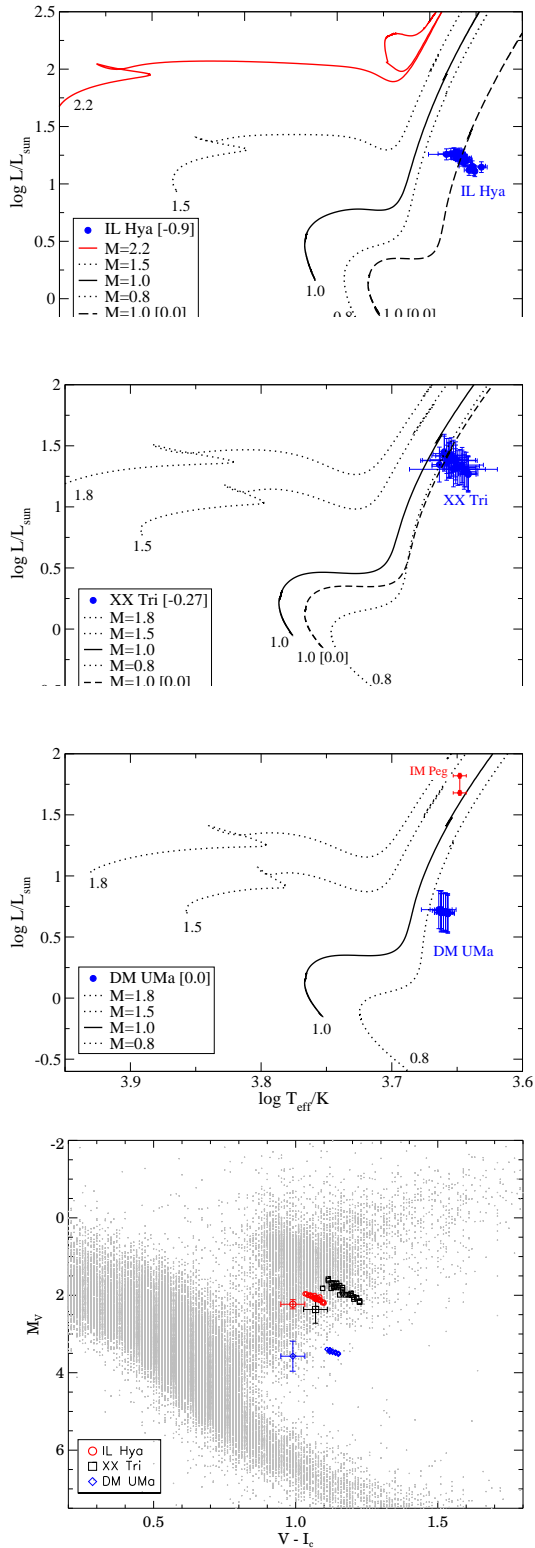}
 \caption{\emph{First three panels from the top:}  log $L/L_{\odot}$/log $T_{\mathrm{eff}}$ of IL~Hya, XX~Tri and DM~UMa, from Table~\ref{temp_lum_il}, Table~\ref{temp_lum_xx} and Table~\ref{temp_lum_dm}, respectively, on theoretical HRDs (Bressan et al. \cite{Bressan}, and post-MS evolution. The dashed tracks are 1~M$_{\sun}$ at solar metallicity, as a comparison to the tracks with non-solar abundances of IL~Hya and XX~Tri. IM~Peg is plotted as a reference star with ([M/H]=0.0, Berdyugina et al. \cite{berd}), see text for details. No stars below 0.8 ~M$_{\sun}$ can have reached the giant phase, as the 0.8~M$_{\sun}$ already requires a theoretical evolutionary age of 25~Gyr. \emph{Bottom:}IL~Hya, XX~Tri and DM~UMa compared to \emph{Hipparcos} stars within 200~pc of the Sun. The measured points for our K giants are all from times of least spottedness.}
         \label{hrd}
   \end{figure}
   
% ----------------------------- Fig. 7.
\begin{figure}
\includegraphics[width=7cm]{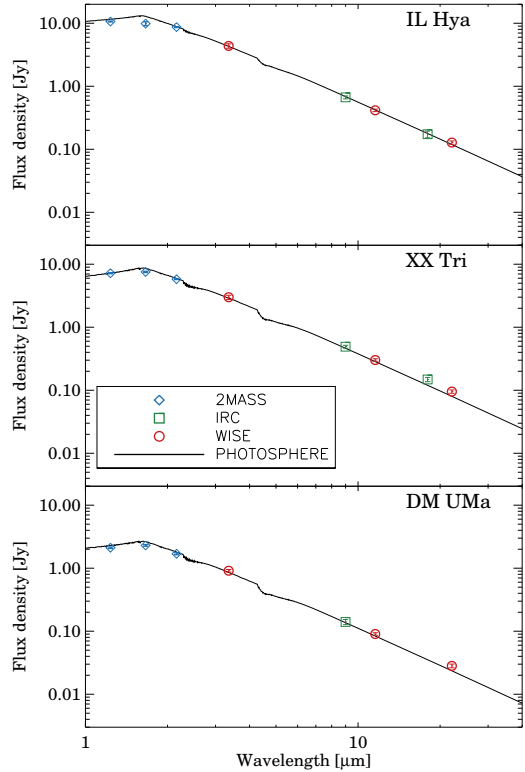} 
\caption {Spectral energy distribution of IL~Hya, XX~Tri, and DM~UMA in the near- and mid-infrared using the ATLAS9 grid of model
atmospheres (Castelli \& Kurucz \cite{kurucz}); see the text for details.
} 
\label{wise} 
\end{figure}

\begin{table}
\caption{Distance, extinction and metallicity of IL~Hya, XX~Tri and DM~UMa.}             
\label{table:8}      
\centering                          
\begin{tabular}{l l l l }       
\hline\hline \noalign{\smallskip}   
      Star  &  distance$^{\it a}$ & $E(B-V)$$^{\it b}$ & [Fe/H]  \\      
\noalign{\smallskip}\hline\noalign{\smallskip}
IL~Hya & 106$\pm5$ & $\approx$0 & $-$0.5$^{\it c}$,$-0.9$$^{\it d}$ \\
XX~Tri & 160$\pm22$ & 0.05 & $-0.27\pm0.03$$^{\it b}$\\
DM~UMa & 139$\pm21$ & $\approx$0 & 0 (assumed) \\
\noalign{\smallskip}\hline
\end{tabular}
\tablefoot{$^{\it a}$Hipparcos (van Leeuwen \cite{hipparcos}), $^{\it b}$this paper, $^{\it c}$Randich et al. (\cite{randich}), $^{\it d}$Weber \& Strassmeier (\cite{michi_klaus}). }
\end{table}

The upper panel of Fig.~\ref{hrd} compares the derived $L$ and $T_{\mathrm{eff}}$ values for IL~Hya with the latest evolutionary tracks from Bressan et al. (\cite{Bressan}). We show post-main-sequence tracks between 2.2~M$_{\sun}$ and 0.8~M$_{\sun}$ for [Fe/H]=$-0.9$, the metallicity suggested for IL~Hya. A solar-mass track is plotted with solar metallicity for comparison. Quite annoyingly, it provides the only obvious ``fit'' to the data, but puts IL~Hya at  an age close to the age of the Universe of 14~Gyr. Lower-mass tracks with any metallicity only fit the observations at stellar ages older than the Universe. From such a comparison, we would conclude a mass of IL~Hya of at most one solar mass for solar metallicity or closer to 0.5~M$_{\sun}$ for the metallicity observed. The most recent minimum mass of 1.194$\pm$0.019 from the orbital elements of Fekel et al. (\cite{fekeletal}), and the inclination of 56\degr$\pm$6\degr \ from the Doppler imaging of Weber \& Strassmeier (\cite{michi_klaus}), suggest a mass of 2.1$\pm$0.6 M$_\odot$ for the K giant. Its error is mostly due to the error in the inclination but not even an inclination of zero degrees, (i.e., when the star is viewed exactly pole-on), would bring it in agreement with the above HRD-based mass, even though this would contradict all rotational-modulation evidence. The same situation is encountered for XX~Tri and DM~UMa (Fig.~\ref{hrd}, second and third panel). Their derived $L/T_{\mathrm{eff}}$ values would render them at impossible masses and ages. 

Morel et al. (\cite{morel1}) show that the effective temperatures of active stars derived from color indices are systematically lower, on average by 175\,K, than the excitation temperature that they used as an estimate of the effective temperature. However, adding this value to our derived temperatures does not affect the results: the observed luminosities are much lower than expected from the stellar models.

The bottom panel of Fig.~\ref{hrd} shows the position of IL~Hya, XX~Tri, and DM~UMa within an observed H-R diagram. It compares our observed absolute $V$ magnitudes and $V-I_C$ indices with \emph{Hipparcos} stars with $\pi\geq5$\,mas and $\pi/\sigma_\pi\geq5$, i.e., stars basically within 200\,pc (van Leeuwen \cite{hipparcos}). The three giants are indicated only with the respective absolute brightness and color indices at light maxima because these are the least spotted states. The comparison shows that our targets are fainter and redder than the bulk of the giant stars observed by \emph{Hipparcos}, even in the least spotted state. A direct comparison of the \emph{Hipparcos} based $M_V$ magnitudes of IL~Hya, XX~Tri, and DM~UMa with our ground-based data also shows them to be fainter and/or redder. 

However, the \emph{Hipparcos} measurements were not exactly simultaneous to any of ours and thus were at random phases of their rotational modulation, which can cause a discrepancy of several tenths of a magnitude compared to the maximum magnitudes.  This explains the apparently fainter \emph{Hipparcos} $M_V$ values. The slight color discrepancy is probably due to the non-trivial determination of $V-I_C$ from $Hp$ and Tycho magnitudes for red stars (see ESA \cite{esa}, Sect. 1.3, Appendix 5). We conclude that the masses and ages of very active stars that are derived from a comparison of the measured luminosities and effective temperatures with evolutionary tracks are unreliable. It is quite possible that the high level of magnetic activity affects the evolution of a star on the giant branch in ways that are still poorly understood.

Another such example is IM~Peg. It served as the guide star for NASA's Gravity-Probe-B mission. IM~Peg is a well-studied K0 active giant primary star of a close binary with strong rotational modulation and activity cycles, just like our three giants. All of its physical parameters, including the radii of the two components, were well-determined because of the needs of that mission (Marsden et al. \cite{marsden}, Berdyugina et al. \cite{berd}), yet its position in the theoretical H-R diagram (Fig.~ \ref{hrd}) does not agree with its derived minimum mass from the orbit and the best-fit inclination from Doppler imaging  (1.8$\pm$0.2~M$_{\sun}$). The star has lower luminosity $\approx$5\% from the measurements than expected from the models. IM~Peg, on the other hand, has only a 0\fm25 difference between the brightest and the faintest maxima, i.e., it has no large, long-term modulation.

\section{Summary and discussion}\label{disc}

We investigated the temperature and luminosity variations of IL~Hya, XX~Tri, and DM~UMa over decade-long timescales. In all three cases it seems that the observed decade-long flux change deduced from multicolor photometry cannot be explained by temperature variations solely due to spots and faculae. For IL~Hya and XX~Tri the flux changes derived from $m_{\rm bol}$ and $T_{\rm eff}$ are markedly different (see Table~\ref{summary_tab}), and the high amplitude long-term change of DM~UMa is not reflected in the variation of the color indices at all.  Table~\ref{summary_tab} is an attempt to summarize the morphology observed and, whenever possible, the numerical values. 

Recently, Tang et al. (\cite{sumin}) reported  the discovery of long-term photometric variability of another three K giants from the DASCH project data. Those objects are very probably also active stars and show variability on a timescale of 10--100 years with amplitudes of about one full magnitude (see Fig.~1 in Tang et al. \cite{sumin}). Interestingly, the DASCH stars are also metal-poor; [Fe/H] is $-0.9$ for two objects and $-0.3$ for one object, comparable to IL~Hya ($-0.9$) and XX~Tri ($-0.27$). 

O'Neal et al. (\cite{oneal}) used TiO-band fits to determine starspot temperatures and areas of XX~Tri and DM~UMa in 2003 and 1998, respectively. For both stars, the average temperature of the visible stellar disk was found to be near 4300\,K. In the case of XX~Tri, it is 
lower by about 100\,K than the average temperature at light curve minimum derived from our $V-I_C$, at least at the time of the observations, but comparable values were found for earlier dates, especially when the amplitude of the rotational modulation was very large. In case of DM~UMa, the average temperature from the TiO-bands is always lower by at least 200\,K compared to our $V-I_C$-based values. Our two Doppler maps of IL~Hya show an average photospheric temperature that is about 300\,K lower than the color index-based calibration. The generally higher photospheric temperatures from $V-I_C$ probably originate from an unremovable facular contribution to the broadband colors, which could be different for spectral lines and the continuum and could be variable in time as well.

\begin{table*}
\caption{Summary of the photometric characteristics, temperatures and luminosities of IL~Hya, XX~Tri and DM~UMa from $V-I_C$ at maximum and minimum brightness.}            
\label{summary_tab}      
\centering                          
\begin{tabular}{l l l l}       
\hline\hline \noalign{\smallskip}   
Parameter   &  IL~Hya & XX~Tri & DM~UMa \\      
\noalign{\smallskip}\hline\noalign{\smallskip}
Cycle from $V$ & trend + 9 + 14 yr  & trend + 6 yr  & trend + 17 yr  \\
Cycle from $B-V$ & \dots & yes & 17 yr cycle \\
Cycle from $U-B$ & \dots & yes & 17 yr cycle \\
Cycle from  $V-I_C$ & trend+7 yr cycle & trend+6.5 yr cycle & small amplitude, uncertain \\
Total brightness change in $V, V-I_C$ [mag] & 0.9, 0.17 (0.25)$^{\it a}$ & 1.05, 0.3 & 0.65, $\approx0.0^{\it b}$ \\
Change of the maxima in $V, V-I_C$ [mag] & 0.5, 0.12 & 0.6, 0.17 & 0.4, $\leq0.05$ \\
Total change in $m_{\rm bol}$ [mag] & 0.55 & 0.7 & 0.22 \\
Change of the maxima in $m_{\rm bol}$ [mag] & 0.36 & 0.46 & 0.09 \\
Temperature change during rotation [K] & 50-150 & 50-200 & 50-100 \\ 
Temperature change of maxima [K] & 210 & 190 & 70 \\
Flux change from $m_{\rm bol}$ [\%] & 39 & 54 & 1 \\
Flux change from $T_{\rm eff}$ [\%] & 20 & 16 & 5 \\
\noalign{\smallskip}\hline
\end{tabular}
\tablefoot{$^{\it a}$ The first two (uncertain) $I$-band data were measured in the Kron-Eggen system (see Sect.~\ref{obs}), which result in a much higher $V-I_C$ amplitude. See text in Sect.~\ref{il_temp}. $^{\it b}$ For DM~UMa, at low $V$ brightness, there is no corresponding $V-I_C$ data.}
\end{table*}

The observed flux change could not originate, e.g., from transits of circumstellar dust in front of the stellar disk because the SED fits to near- and mid-infrared data of our targets do not show any excess (Fig.~\ref{wise}).  
For each star we selected the model from the ATLAS9 grid of model atmospheres (Castelli \& Kurucz 2003) that has the closest metallicity
and surface gravity, and we interpolated between the two closest models in effective temperature. The effective temperatures of our targets were
taken from the corresponding table (Tables~\ref{temp_lum_il}, \ref{temp_lum_xx}, and  \ref{temp_lum_dm}), and always using $T_{\rm eff}$ obtained in the last epoch at maximum brightness. The model spectrum was then scaled to the 2MASS K-band and WISE1 band photometric points. We note that the WISE 4.6$\mu$m measurements were not plotted because of the well-known systematic overestimation effect of the fluxes of bright sources in this band (Section 6.3 of the Explanatory Supplement to the WISE All-Sky Data Release Products).

The longest period pulsating stars with similar light variation amplitudes, colors, and luminosities compared to our objects are among the semiregular (SR) d-type variables. Giridhar et al. (\cite{giridhar1}, \cite{giridhar2}, \cite{giridhar3}) studied a few metal-poor  SRd stars in detail. However, those pulsating stars have variability timescales only up to a few hundred days, while the long-term changes in our objects are decades long. The observed possible radius change in IL~Hya also contradict pulsation: we find the largest radius at the brightest state of the star, while in case of pulsation the increasing brightness goes together with decreasing radius.

Luminosity changes with nearly constant effective temperature could in fact mean radius changes. In the case of IL~Hya this is suggested both by the radii from $v\sin i$ measurements and those we derive from bolometric luminosities at different epochs, although with marginal significance (see Table~\ref{temp_lum_il}). The timescale of just a few decades for such radius changes is much too short to be attributable to normal evolutionary effects on the red-giant branch (RGB). Yet the thermal pulses on the asymptotic giant branch (AGB) occur on a decadal timescale. 
However, the evolutionary position of IL~Hya and other red giant RS~CVn stars is thought to be on the first ascent of the RGB, which rules out an AGB scenario as well. The timescales of the different evolutionary stages after the main sequence is well described by Sackmann et al. (\cite{sackmann}) for the Sun, which gives a good hint on our stars' evolution as well.

Measuring the solar radius dates back several decades with contradictory results. It is not clear whether the solar radius is constant or variable. If it is variable, is there any connection between the radius change and solar activity?  From full-disk solar images at 6723\,\AA\, Chapman et al. (\cite{chapman}) deduced a small radius change in the Sun in phase with the 11 yr spot cycle. The radius is larger at stronger activity, although with small significance.  Earlier work by Laclare et al. (\cite{laclare}) showed the opposite behavior; the solar radius was smaller when more spots were present, at least for more than 1.5~solar cycles between 1978 and 1994. Using the same dataset as Laclare et al. (\cite{laclare}), but extended to the year 2000, Qu~\&~Xie (\cite{qu}) came to the same conclusion; the Sun shrinks when the activity is high, probably due to cyclic variation of the magnetic pressure in the solar convection zone. On the basis of the magnetic virial theorem, Stothers (\cite{stothers}) suggests that the solar radius changes by about 0.02\% with the 11 yr cycle, and decreases around the maximum activity.  Mullan et al. (\cite{mullan}) found different radius changes in a solar model during different period lengths, and arrived at the conclusion that, \emph {"According to [their] models, the phase of the luminosity variation is such that the Sun is predicted to be least luminous when the global magnetic field strength is strongest".} Assuming that the spot cycles of our active giants are of similar origin to the solar cycles, this result is in line with our observation of a smaller radius of IL~Hya when it is fainter, i.e.,  when the star shows higher activity with more spots (see Table~\ref{radius_il}). 

At present, we cannot provide a conclusive explanation for the origin of the one-magnitude brightness variations in our three active giant stars, but a cyclic radius change is among the possible causes. In addition to the starspot-related effective temperature changes, the global magnetic field could be the agent that changes the radius of the star that then causes a color-neutral brightness variation.  Magnetic activity cannot yet be taken into account when calculating evolutionary tracks, but have the potential to also explain why these stars have much lower luminosities than expected from evolution models. It appears that the global magnetic fields are behind the long-term changes in IL~Hya, XX~Tri, and DM~UMa, and perhaps in other stars as well.

\begin{acknowledgements}
We are grateful to K. Stepien who raised the questions of the huge brightness variations of active stars back in 2006 during the IAU GA in Prague. Thanks are also due to M. Weber for determining the [Fe/H] value for XX~Tri, and to  A. Lanza and S. Messina for helpful discussions. We thank an anonymous referee for constructive suggestions.
This work has been supported by the Hungarian Science Research Program OTKA K-81421, OTKA-109276, the Lend\"ulet-2012 Young Researchers' Programs of the Hungarian Academy of Sciences.
This publication makes use of data products from the Two Micron All Sky Survey,
which is a joint project of the University of Massachusetts and the Infrared
Processing and Analysis Center/California Institute of Technology, funded by the
National Aeronautics and Space Administration and the National Science
Foundation. This publication makes use of data products from the Wide-field Infrared Survey
Explorer, which is a joint project of the University of California, Los Angeles,
and the Jet Propulsion Laboratory/California Institute of Technology, funded by
the National Aeronautics and Space Administration.

\end{acknowledgements}

￼￼\appendix

\section{The dependence of the measured $v\,{\rm sin}\,i$ on the spottedness of a star}\label{appA}

We tested the dependency of the measured $v\sin{i}$ on the spot coverage of IL~Hya. For this, we have assumed different spot distributions with both polar caps and equatorial belts simultaneously. The corresponding line profiles with a given $v\sin{i}$ were calculated using the forward version of our Doppler imaging code \texttt{TempMapForward} (Rice \& Strassmeier, \cite{rice_strass_2000}). Finally, we measured the FWHM of the lines to derive $v\sin{i}$. The stellar parameters of IL~Hya used were those given in Section~\ref{section_di} and in Weber \& Strassmeier (\cite{michi_klaus}), and were kept constant over the whole course of the test, only the polar and equatorial spot-coverage was gradually changed. These tests were carried out for polar caps of different areas with fixed equatorial belts, and \emph{vice versa}. We note, that these spot configurations are symmetric and do not result in rotational modulation. We also note, that microturbulence affects EWs and therefore the measured FWHMs, but this also did not change during the tests.

The results clearly indicate that a shrinking polar cap with a fixed equatorial belt results in significantly lower $v\sin{i}$, while a shrinking belt with a fixed cap produces constant, or slightly higher $v\sin{i}$. The shrinking huge polar cap yields slightly higher $v\sin{i}$ at first, and then significantly lower $v\sin{i}$ later on. 

   \begin{figure}[]
   \centering
   \includegraphics[width=7cm]{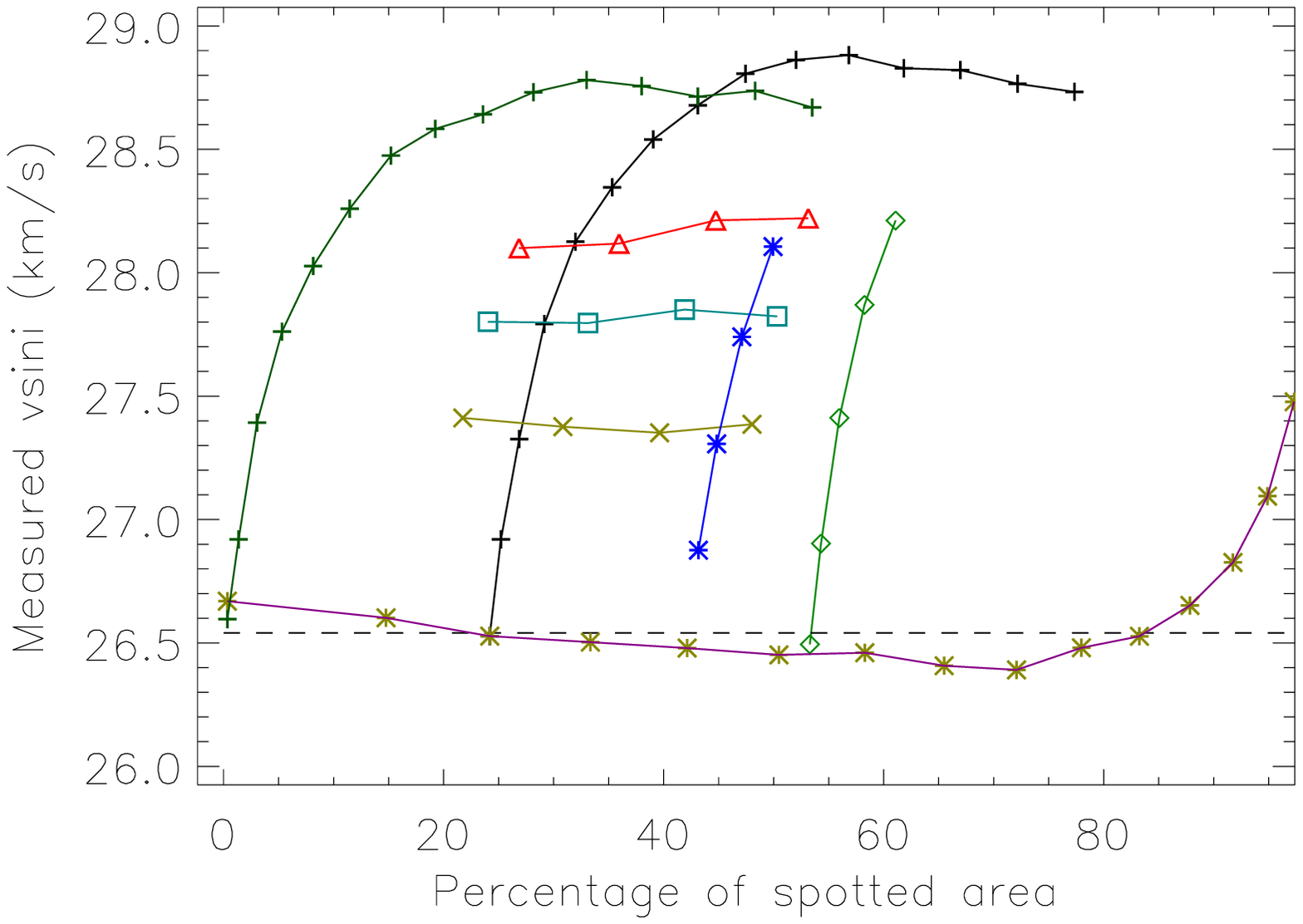}
   \includegraphics[width=2cm]{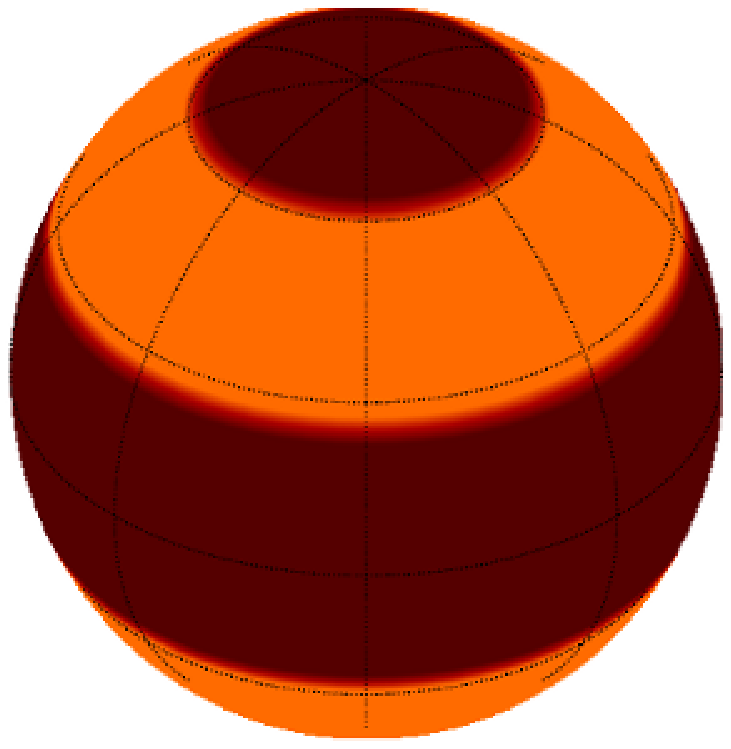}\hspace{5mm} \includegraphics[width=2cm]{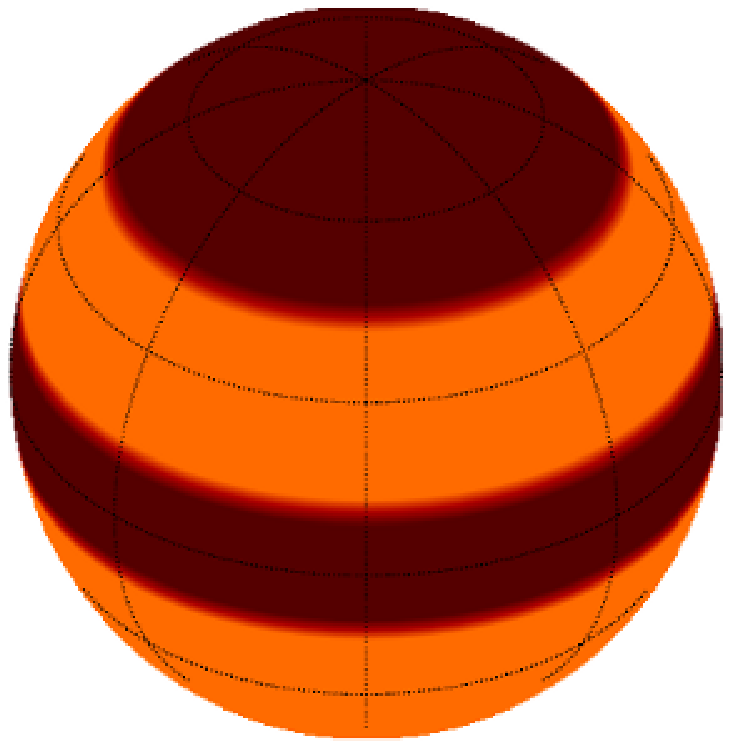}
   \caption{Dependency test of measured $v\,{\rm sin}\,i$ on spot coverage. Shrinking polar caps from $30\degr$ radius by $5\degr$ steps with a constant equatorial belt between $-30\degr$ and $+30\degr$ and $-25\degr$ and $+25\degr$ latitudes are represented by diamonds and stars, respectively; the effect of a huge shrinking polar cap from $45\degr$ radius by $5\degr$ steps with a thin equatorial belt between $-10\degr$ and $+10\degr$, and without equatorial belt, are represented by plusses (vertical structures from the right to the left). The starting  $v\,{\rm sin}\,i$ of 26.5 is represented by a dashed line, around it is seen the effect of an equatorial belt without a polar cap marked with stars; crosses, squares and triangles denote shrinking equatorial belts from $-30\degr$ and $+30\degr$ with $5\degr$ steps and with constant radii of polar caps of $15\degr, 20\degr$ and $25\degr$, respectively (horizontal structures upwards). Below: two representative starting configurations are given to visualize the spots. }
         \label{vsini_test}
   \end{figure}
  
\end{document}